\documentclass[epj]{svjour}

\usepackage{latexsym}
\usepackage{graphicx}
\usepackage{subfigure}
\usepackage[dvips]{color}
\usepackage{enumitem}

\begin{document}

\title{Density-temperature scaling of the fragility in a model glass-former}
\author{Shiladitya Sengupta $^{1,}$\thanks{email:shiladityas@tifrh.res.in}
Thomas B. Schr\o der $^{2,}$\thanks{email:tbs@ruc.dk}
\and Srikanth Sastry  $^{1,3,}$\thanks{email: sastry@tifrh.res.in}  
}                     

\titlerunning{DT Scaling of fragility}
\authorrunning{Sengupta, Schr\o der, Sastry}
\institute{
TIFR Centre for Interdisciplinary Sciences, Tata Institute of Fundamental Research, 21 Brundavan Colony, Narsingi, Hyderabad 500075, India.
\and
DNRF Centre “Glass and Time”, IMFUFA, Department of Science, Systems and Models, Roskilde University, Postbox 260, DK-4000 Roskilde, Denmark.
\and
Theoretical Sciences Unit, Jawaharlal Nehru Centre for Advanced Scientific Research, Jakkur Campus, Bangalore 560064, India. 
}

\date{\today} 

\abstract{
Dynamical quantities such as the diffusion coefficient and relaxation times for some glass-formers may depend on density and temperature through a specific combination, rather than independently, allowing the representation of data over ranges of density and temperature as a function of a single scaling variable. Such a scaling, referred to as density - temperature (DT) scaling, is exact for liquids with inverse power law (IPL) interactions but has also been found to be approximately valid in many non-IPL liquids. We have analyzed the consequences of DT scaling on the density dependence of the fragility in a model glass-former. We find the density dependence of kinetic fragility to be weak, and show that it can be understood in terms of DT scaling and deviations of DT scaling at low densities. We also show that the Adam-Gibbs relation exhibits  DT scaling  and the scaling exponent computed from the density dependence of the activation free energy in the Adam-Gibbs relation, is consistent with the exponent values obtained by other means. 
} 

\maketitle
\section{Introduction}\label{intro}
Glass forming liquids exhibit significant diversity in their dynamical behaviour as the glass transition is approached. Any concept that helps to organize and simplify the description of dynamics is therefore very useful, especially in the absence of a satisfactory general theoretical framework to describe all phenomenology in full detail. Fragility, which measures how steeply the dynamical quantities (viscosity, relaxation time and inverse diffusivity) increase as a liquid is cooled, is a material parameter that has proved effective to  compare the slowdown of dynamics in different liquids. Understanding the origin of the fragility of glass-formers therefore has become an important part of explaining glassy behaviour. Similarly, recent analyses reveal that for many liquids density and temperature are not independent parameters controlling dynamics but can be combined into a \emph{single} thermodynamic control parameter. So dynamical quantities like diffusivity and relaxation time become function of a specific combination of density and temperature which is denoted here as the ``Density-Temperature Scaling'' (DT) and will be discussed in more detail subsequently. Note that such a scaling relation indicates that in some liquids the number of independent thermodynamic control parameters required to describe dynamics can be reduced and hence implies the interesting possibility that some liquids may be \emph{inherently} simpler than others. Specifically, if the density-temperature scaling were exact, then the diffusivities (relaxation times) at different densities and temperatures could be collapsed on a single master curve with only one independent variable. Hence the kinetic fragility would be independent of density. As we discuss subsequently, DT scaling is exact for a class of model liquids in which the interparticle interactions are described by purely repulsive inverse power law (IPL) potentials : $u(r_{ij}) = \epsilon \left(\frac{\sigma}{r_{ij}} \right)^{n}$. For a \emph{non}-IPL liquid, there is no {\it a priori} reason to expect DT scaling.  In the presence of such scaling, the aim of the present study is to understand the consequences of the density temperature scaling on the fragility in a \emph{non}-IPL glass-forming liquid, namely, the Kob-Andersen binary mixture. 

The power law nature of potentials in IPL systems has a number of simplifying consequences. First, the interaction potential $u(r_{ij})$ being a homogeneous function has the consequence that by choosing the unit of length to be $l_{0} = \rho^{-1/D}$ in $D$ dimensional space, and the unit of time to be $\tau = l_0 \left(m l_{0}^{n} \over \epsilon \sigma^{n}\right)^{1/2}$ (corresponding to a unit of energy of $\epsilon_0 = \epsilon \sigma^n/l_0^n$), the equations of motion are invariant for a change in density\cite{pap:invpower-hiwatari}. Second, when an IPL liquid is kept at a constant temperature $T$, in the system of units ($l_{0}, \epsilon_{0}, m$), the canonical probability $\beta H$ of observing a microstate in the phase space depends only on the reduced temperature  $T^{*}=\frac{k_{B}T}{\epsilon\sigma^{n}\rho^{n/D}}$ ($H$ is the Hamiltonian, $\beta=1 / k_{B}T$).  Thus the probability distribution $\frac{e^{-\beta H}}{Z}$, the partition function $Z(N,\rho,T)$ and the ensemble average $\langle g(\vec{p}^{N},\vec{r}^{N})\rangle$  of a phase-space observable, in the reduced system of units ($l_{0}, \epsilon_{0}, m$) become functions of only the reduced temperature $T^{*}$ in which the density and the temperature appear through the specific combination $\frac{\rho^{n/D}}{ T}$. Thus the thermodynamic and dynamic properties of an IPL liquid at different densities and temperatures having the same value of $\frac{\rho^{n/D}}{ T}$ are the same. This property is denoted here as the ``Density-Temperature Scaling'' (DT) \cite{pap:invpower-hiwatari,pap:invpower,pap:invpower-Hooveretal}. A third relevant property of an IPL model as a consequence of the interparticle interaction potential $u(r_{ij})$ being a homogeneous function of $r_{ij}$ is that the instantaneous potential energy ($U$) is proportional to the instantaneous virial ($W$): $U = \frac{n}{D} W$.

Recent analyses of liquid state properties have \emph{empirically} shown \cite{pap:Bohling-etal,pap:Tolle,pap:Dreyfus-etal-1,pap:Dreyfus-etal-2,pap:roland-etal-DT-review,pap:Alba-simionesco-DTscaling-twoforms,pap:scaling-4,pap:scaling-5,pap:coslovich-roland-KA-DTscaling-tau} 
that many \emph{non}-IPL liquids - {\it e.g.} liquids interacting \emph{via} Van der Waals interactions  - also obey  density-temperature scaling. This implies that, even in \emph{non}-IPL liquids, relaxation times and diffusivities depend on densities and temperatures through a specific combination, rather than independently. Over a broad range of densities (pressures) and temperatures, this combination was generally found to be $\frac{\rho^{\gamma}}{T}$, {\it i.e.} $\tau = f(\frac{\rho^{\gamma}}{T})$ where the function $f$ is in general unknown and material-specific, although other forms have also been suggested \cite{pap:Alba-simionesco-DTscaling-twoforms,pap:Alba-Simionesco-Tarjusscaling}. The exponent $\gamma$ was found empirically by collapse of relaxation time data at different densities (pressures) and temperatures on a single master curve.

The remarkable data collapse for relaxation times using the functional form $\tau(\rho,T) = f(\rho^{\gamma} / T)$ naturally suggests that inter-particle interactions of \emph{non}-IPL liquids showing DT scaling can be approximated by an \emph{equivalent} IPL model. Dyre, Schr\o der and co-workers  have analyzed this idea for the Lennard-Jones interaction and have shown that the interaction potential and the fluctuations in potential energy and virial are captured to a good approximation by a potential consisting of  an inverse power law  and a linear term \cite{pap:scaling-2}.  They have further identified a class of \emph{non}-IPL liquids for which the properties of IPL systems discussed previously hold to a very good approximation \cite{pap:scaling-4,pap:scaling-5,pap:scaling-2,pap:scaling-1,pap:scaling-3}. In particular, (1) these liquids show strong correlation between instantaneous values of the potential energy ($U$) and the virial ($W$). Hence they are denoted as ``strongly correlating liquids''. We note that by definition, the IPL systems show exact correlation. Further, it was shown that the scaling exponent $\gamma$ in these \emph{non}-IPL liquids can be \emph{computed} from the virial-potential energy correlation {\it i.e.} the scaling exponent is not an empirical parameter but can be \emph{predicted} from thermodynamics. (2) Strongly correlating liquids show an \emph{approximate} property called ``isomorphism'' :  If any configuration at a state point $(\rho_{1}, T_{1})$ and another at a state point $(\rho_{2},T_{2})$ have the same reduced coordinates in appropriate units (discussed later), then their Boltzmann weights are proportional : $e^{-U_{1} / k_{B}T_{1}} = C_{12} e^{-U_{2}/k_{B}T_{2}}$ where the constant of proportionality $C_{12}$ depends on the state points but is independent of microscopic configurations. Again for IPL systems, the isomorphism is exact, with $C_{12}=1$.  
An isomorph is then a curve in the phase diagram where all pairs of state points are isomorphic according to the definition above. The isomorph theory predicts that the structure and dynamics is invariant on an isomorph (provided the proper reduced units are used, see below). Also the excess entropy (compared to the ideal gas at the same density and temperature), the configurational entropy, and the isochoric heat capacity are predicted to be invariant on an isomorph. From a practical point of view, an important isomorph invariant is $h(\rho)/T$ where $h(\rho)$ is a function that depends on the system in question. For atomic systems where the interaction potential is a sum of power-laws, e.g., the Lennard-Jones potential, there exists an analytical expression for $h(\rho)$, and consequently the invariance of $h(\rho)/T$ can be used to identify isomorphs spanning large density changes \cite{pap:Bohling-etal}. For smaller density changes $h(\rho)$ can be well approximated by $\rho^\gamma$ where $\gamma$ can be calculated independently from the equilibrium fluctuations of the virial and the potential energy (see section 3.6 below). Since $\tau$ is also an isomorph invariant, this in turn explains the observed power law density scaling $\tau = f(\rho^\gamma/T)$, but also its breakdown at large density variation where $\tau = f(h(\rho)/T)$ should be used \cite{pap:Bohling-etal}. For the density changes considered in this paper it is sufficient to use $h(\rho)=\rho^\gamma$, and to simplify the analysis we will do so.
To summarize, isomorphic state points have the same dynamics, which explains the density-temperature scaling in some \emph{non}-IPL liquids. However, we note that, two things complicate this picture. First, the values of $\gamma$ computed from  virial-potential energy correlation depend on the state point, but this problem can be tackled by using a general form of $h(\rho)$. Second, Tarjus and coworkers have come up with a counter example (WCA version of the KA liquid) which show strong correlation but not density-temperature scaling \cite{pap:BT-PRL,pap:BT-2011} although it is a subject of debate \cite{pap:Toxvaerd-Dyre}.

Here, we study the consequences of the DT scaling on the density  dependence of the kinetic and the thermodynamic fragilities (see Sec. \ref{sec:defn} for definitions) of a model glass-former \emph{viz.} the  Kob-Andersen (KA) model \cite{kob} in three dimensions. If the DT scaling were \emph{exact}, then the kinetic fragility would be independent of density while the thermodynamic fragility would show power law dependence on density.  We verify that both diffusivity and relaxation time show impressive DT scaling although there are deviations at low densities.  We find that the density dependence of kinetic fragility is weak and show that  it can be understood in terms of DT scaling and deviations of DT scaling at low densities.  We also show that the Adam-Gibbs relation exhibits DT scaling at high densities and the scaling exponent computed from the density dependence of the activation free energy in the Adam-Gibbs relation is consistent with the exponent values computed by other means.


\begin{table}[!]
\caption{\label{tab:statepoints}  Range of density, temperature chosen}
\begin{tabular}{cc}
Density / $\sigma_{AA}^{-3}$ & Temperature range / $\epsilon_{AA}k_{B}^{-1}$\\
\hline
1.10 & 0.28 - 2.00\\
1.15 & 0.34 - 2.00\\
1.20 & 0.435 - 2.00\\
1.25 & 0.52 - 2.00\\
1.35 & 0.77 - 2.00\\
\hline
\end{tabular}
\end{table}

\section{Simulation details}\label{sec:simudetail}
We have studied the Kob-Andersen (KA) model \cite{kob} which is a binary mixture ($A_{80}B_{20}$) of Lennard Jones particles. The interaction potential is truncated such that both the potential and the force go to zero smoothly at a cutoff distance and is given by : 
\begin{eqnarray}
V_{\alpha\beta}(r)&=&4 \epsilon_{\alpha\beta} \left[ \left( \frac{\sigma_{\alpha\beta}}{r} \right)^{12} - \left( \frac{\sigma_{\alpha\beta}}{r} \right)^{6} \right] \nonumber\\
&+& 4 \epsilon_{\alpha\beta}\left[c_{0 \alpha\beta} + c_{2 \alpha\beta}\left(\frac{r}{\sigma_{\alpha\beta}}\right)^{2}\right], r_{\alpha\beta} < \mbox{cutoff} \nonumber\\
&=& 0, r_{\alpha\beta} > \mbox{cutoff} \nonumber
\label{eqn:4DKAmodel}
\end{eqnarray}
where $\alpha,\beta \in \{A,B\}$.  The units of mass, length, energy and time scales are $m_{AA}, \sigma_{AA},\epsilon_{AA}$ and
$\sqrt{\frac{\sigma_{AA}^{2}m_{AA}}{\epsilon_{AA}}}$ respectively. We denote this system of units as the \emph{conventional reduced units} to distinguish from the system of reduced units appropriate for density-temperature scaling defined in section \ref{sec:DTSunit}. In the conventional units, $\epsilon_{AB}=1.5$, $\epsilon_{BB}=0.5$, $\sigma_{AB}=0.80$, $\sigma_{BB}=0.88$. The interaction potential was cutoff at $2.5 \sigma_{\alpha\beta}$.
The correction terms $c_{0\alpha\beta}, c_{2\alpha\beta}$ are obtained from:
\begin{equation}
V_{\alpha\beta}(r_{c\alpha\beta}) = 0, \left(\frac{d V_{\alpha\beta}}{d r}\right)_{r_{c\alpha\beta}} = 0\nonumber
\end{equation}

Molecular dynamics (MD) simulations were done in a cubic box with periodic boundary conditions in the constant number, volume and temperature  (NVT) ensemble at five number densities ($\rho=1.10, 1.15, 1.20,  1.25, 1.35$). The integration time step was in the range $dt = 0.001-0.005$. Temperatures were kept constant using the Brown and Clarke implementation of the Hoover dynamics \cite{pap:BC}. System size was $N=1000, N_{A}=800$ ($N =$ total number of particles, $N_{A}=$ number of particles of type $A$). The temperature ranges simulated for different densities are shown in Table \ref{tab:statepoints}. For all state points, three to five independent samples with run lengths $>100 \tau_{\alpha}$ ($\tau_{\alpha}$ is the $\alpha $- relaxation time) were analyzed.

\section{Definitions}\label{sec:defn}
Here we briefly summarize the definitions of various  quantities computed and the methods used to compute them.

\subsection{Diffusivities  and $\alpha$ relaxation times}

Diffusivities ($D_{A}$) are obtained from the mean squared displacement (MSD) of the $A$ type particles.

Relaxation times are obtained from the decay of the overlap function $q(t)$ using the definition $q(t=\tau_{\alpha},T)/N =1/e$. The overlap function is a two-point time correlation function of local density $\rho(\vec{r},t)$ which has been used in many recent studies of slow relaxation \cite{pap:4pt-CD,pap:Ovlap-Glotzer-etal,pap:Lacevic,pap:Ovlap-Donati-etal,pap:Karmakar-PNAS}. It is defined as: 
\begin{eqnarray}
<q(t)> &\equiv& <\int d\vec{r}\rho(\vec{r},t_{0})\rho(\vec{r},t+t_{0})>\nonumber\\
&=& <\sum_{i=1}^{N} \sum_{j=1}^{N} \delta (\vec{r}_{j}(t_{0}) - \vec{r}_{i}(t+t_{0}))>\nonumber\\
&=& <\sum_{i=1}^{N} \delta (\vec{r}_{i}(t_{0}) - \vec{r}_{i}(t+t_{0}))> \nonumber\\
&  & + <\sum_{i} \sum_{j\neq i} \delta (\vec{r}_{i}(t_{0}) - \vec{r}_{j}(t+t_{0}))>
\label{eqn:qtdef}
\end{eqnarray}
Here the averaging over time origins $t_{0}$ is implied. In our work, we consider only the self part of the total overlap function ({\it i.e.} neglecting the $i\ne j$ terms in the double summation), which was shown \cite{pap:Lacevic} to be a good approximation to the full overlap function. Thus we have used the definition :  
\begin{eqnarray}
<q(t)>&\approx& <\sum_{i=1}^{N} \delta (\vec{r}_{i}(t_{0}) - \vec{r}_{i}(t+t_{0}))>
\end{eqnarray}
Because we wish to consider configurations that differ only by fast thermal motions as being the same, and for computational convenience, the $\delta$ function is approximated by a window function $w(x)$ which defines the condition of ``overlap'' between two particle positions separated by a time interval $t$: 
\begin{eqnarray}
<q(t)> &\approx& <\sum_{i=1}^{N} w(|\vec{r}_{i}(t_{0}) - \vec{r}_{i}(t_{0}+t)|)>\nonumber\\
w(x) &=& 1, x \leq a \textnormal{  implying ``overlap''}\nonumber\\
     &=& 0 \mbox{   otherwise   }
\end{eqnarray}
The definition of the overlap function thus depends on the choice of the cutoff parameter $a$, which is chosen such that $a^{2}$ is comparable to the value of the MSD in the plateau regime. We choose $a$ to be $0.3$ at all densities.

\subsection{The configurational entropy}
The configurational entropy ($S_{c}$) (per particle), the measure of the number of distinct local energy minima, is calculated \cite{pap:Sc-Sastry} by subtracting the ‘‘vibrational’’ component from the total entropy of the system :
\begin{equation}
S_{c͑}(T) =  S_{total}(T) - S_{vib} (T)\label{eqn:Sc}
\end{equation}
The total entropy of the liquid is obtained via thermodynamic integration from the ideal gas limit. The vibrational entropy is calculated by making a harmonic approximation to the potential energy about a given local minimum \cite{pap:Sc-Sastry,pap:Sc-Sastry-JPCM,pap:PEL-Sciortino,pap:PEL-Heuer} following the procedure described in \cite{pap:Sc-Sastry,pap:Sc-Sastry-JPCM}.

\subsection{Fragility}
Fragility \cite{fragility_angell,pap:Laughlin-Uhlmann-TgbyT} is a material parameter which quantifies how rapidly dynamical quantities (viscosity, relaxation times and inverse diffusivities) rise as  temperature of glass forming liquids are decreased. Fragility has been defined in a variety of ways. We denote the fragilities defined from dynamical quantities as {\it kinetic fragilities}, to distinguish from {\it thermodynamic fragility} defined later. Two of the popular definitions of the kinetic fragility are (i) the ``steepness index'' $m$ defined from the so-called Angell plot as the slope ($m$) of logarithm of the viscosity ($\eta$) or relaxation time ($\tau$) at the laboratory glass transition temperature $T=T_{g}$, with respect to the scaled inverse temperature $T_g/T$ : $m = \left( \frac{d \log \tau}{d (\frac{T_{g}}{T})}\right )_{T=T_{g}}$ and (ii) the fragility defined using Vogel-Fulcher-Tammann (VFT) fits to the temperature dependence of dynamical quantities : 
\begin{equation}
\tau(T) = \tau_{0} \exp\left[\frac{1}{K_{VFT}(\frac{T}{T_{VFT}}-1)} \right] \label{eqn:kinfr2}
\end{equation} 
where $K_{VFT}$ is the \emph{kinetic fragility}  and $T_{VFT}$ is the VFT divergence temperature. In the present study we have used $K_{VFT}$ as the measure of the fragility.

In many systems including the Kob-Andersen model \cite{pap:AG-Sastry}, dynamics can be related to thermodynamics {\it via} the Adam Gibbs (AG) relation \cite{AdamGibbs} between the relaxation time and the configurational entropy ($S_{c}$):
\begin{equation}
\tau(T) = \tau_{0} \exp(\frac{A}{TS_{c}})
\label{eqn:AG}
\end{equation} 
where $\tau_{0}$ and $A$ are material-specific coefficients. Previous work \cite{pap:AG-Sastry} has shown that the configurational entropy can be computed from the properties of the potential energy landscape and thus fragility can be related to thermodynamic quantities {\it via} the AG relation. If the temperature dependence of $S_c$ is given by 
\begin{equation}
TS_{c} = K_{T}\left(\frac{T}{T_{K}}-1\right)
\label{eqn:def-KAG}
\end{equation}
the Adam-Gibbs relation [Eqn. \ref{eqn:AG}] yields the VFT relation provided $T_{VFT} = T_{K}$. We denote $K_T$ as the \emph{thermodynamic fragility}. It is related to the kinetic fragility as : 
\begin{equation}
K_{VFT} = K_T/A  
\label{eqn:KT-KVFT}
\end{equation}
when the Adam-Gibbs relation holds.

\subsection{Reduced units appropriate for the density-temperature scaling}\label{sec:DTSunit}
Let us consider $N$ particles in a box of volume $V$ at constant temperature $T$ and density $\rho$. Reduced unit system appropriate for density-temperature scaling is defined by choosing $l_{0}=(V/N)^{1/3}=\rho^{-1/3}$ as the unit of length, the mass of one atom $m$ as the unit of mass and $t_{0} = \rho^{-1/3}(k_{B}T/m)^{-1/2}$ as the unit of time \cite{pap:Schroeder-etal-Units,pap:Fragiadakis-Roland}. The conversion of dynamical quantities from conventional to this reduced unit system (denoted by $^{*}$) is given by:
\begin{eqnarray}
\tau^{*} &=& \tau / t_{0} = \rho^{1/3} (k_{B}T/m)^{1/2} \tau \nonumber\\
D^{*} &=& D / (l_{0}^{2} t_{0}^{-1}) = \rho^{1/3} (k_{B}T/m)^{-1/2} D
\end{eqnarray}
We have checked that in the studied range of densities and temperatures, the conversion factors change maximum by a factor of 3 {i.e.} the conversion factors are $\mathcal{O}{(1)}$. Hence if the reduced quantities show denstiy-temperature scaling then so does the \emph{bare} relaxation time ($\tau$) and  diffusivities ($D$).

\subsection{Pressure-energy correlation}
The correlation between instantaneous potential energy ($U$) and virial $W$ \cite{pap:scaling-4,pap:scaling-5} is measured in terms of the coefficient
\begin{equation}
R = \frac{\langle\Delta W \Delta U\rangle}{\sqrt{\langle(\Delta W)^{2}\rangle 
\langle (\Delta U)^{2}) \rangle}} 
\end{equation}
where $\Delta U = U - \langle U \rangle, \Delta W = W - \langle W \rangle$ represent instantaneous fluctuations about mean values of potential energy and virial respectively. A liquid is considered to be strongly correlating at a given state point ($\rho,T$) if $R \ge 0.9$.

\subsection{Scaling exponent $\gamma$}
The exponent $\gamma(\rho,T)$ in the functional form $\tau(\rho,T) = f(\rho^{\gamma} / T)$ can be computed from the fluctuations in potential energy and virial as : 
\begin{eqnarray}
\gamma &=& \frac{\langle \Delta W \Delta U \rangle} {\langle (\Delta U)^{2} \rangle}\nonumber
\end{eqnarray}

If the density-temperature scaling of the form $\tau(\rho,T) = f(\rho^{\gamma} / T)$ holds, the density dependent activation energy parameters $A(\rho)$ in Adam Gibbs relation [Eqn. \ref{eqn:AG}] and $E_{0}(\rho)$ in Arrhenius law $\tau(\rho,T)=\tau(\rho,\infty)\exp(E_{0}(\rho)/T)$ are expected to have the density dependence of the form:
\begin{eqnarray}
A(\rho) &\sim& \rho^{\gamma}\nonumber\\
E_{0}(\rho) &\sim& \rho^{\gamma}\nonumber
\end{eqnarray}
which provides another way to compute $\gamma$.

\begin{figure*}[t!]
\begin{center}
\vspace{3mm}
\includegraphics[scale=0.4]{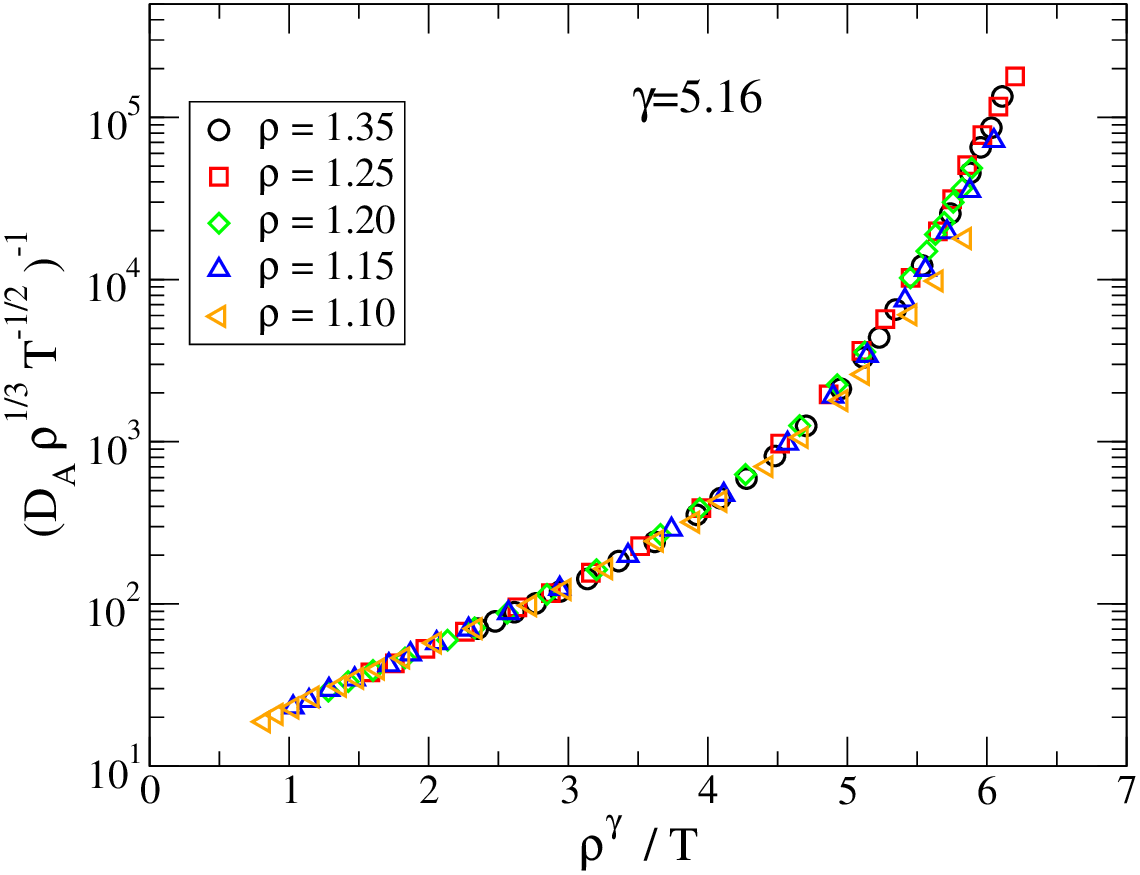} 
\hspace{3mm}
\includegraphics[scale=0.4]{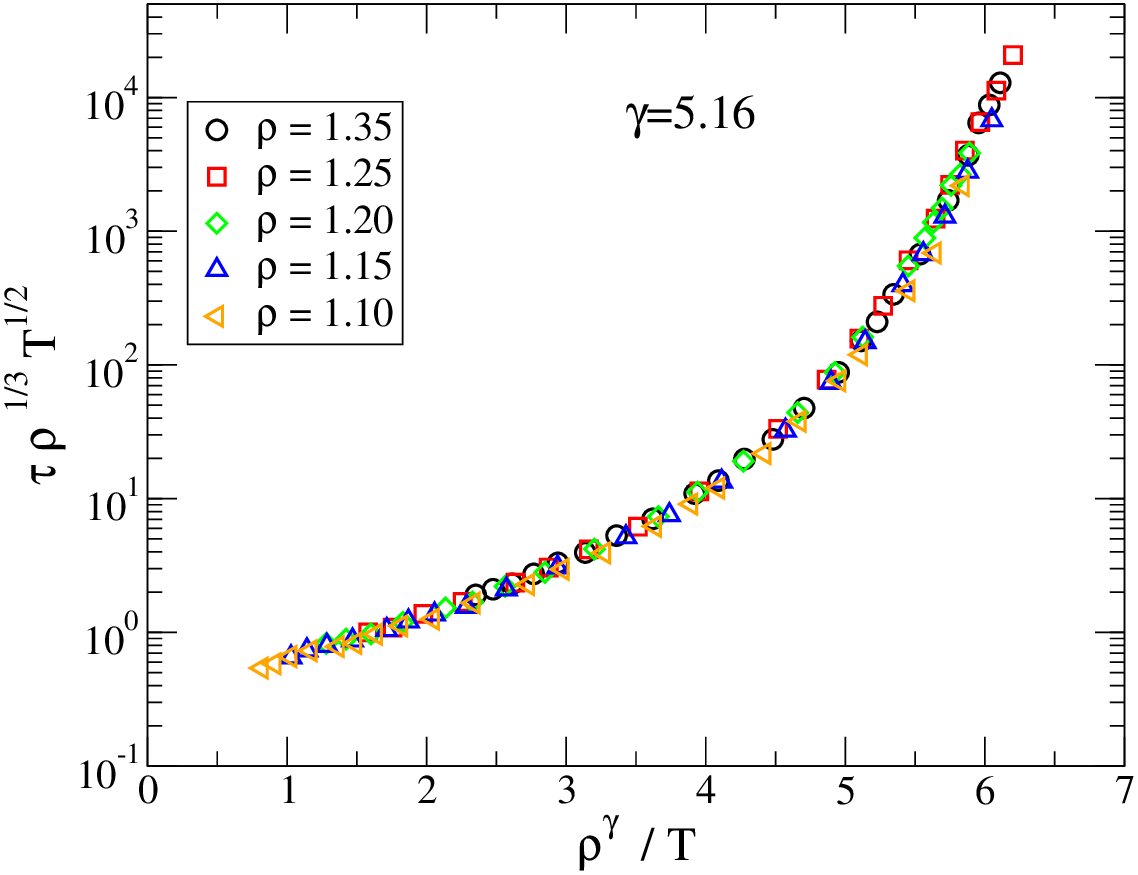} 
\\\vspace{8mm}
\includegraphics[scale=0.4]{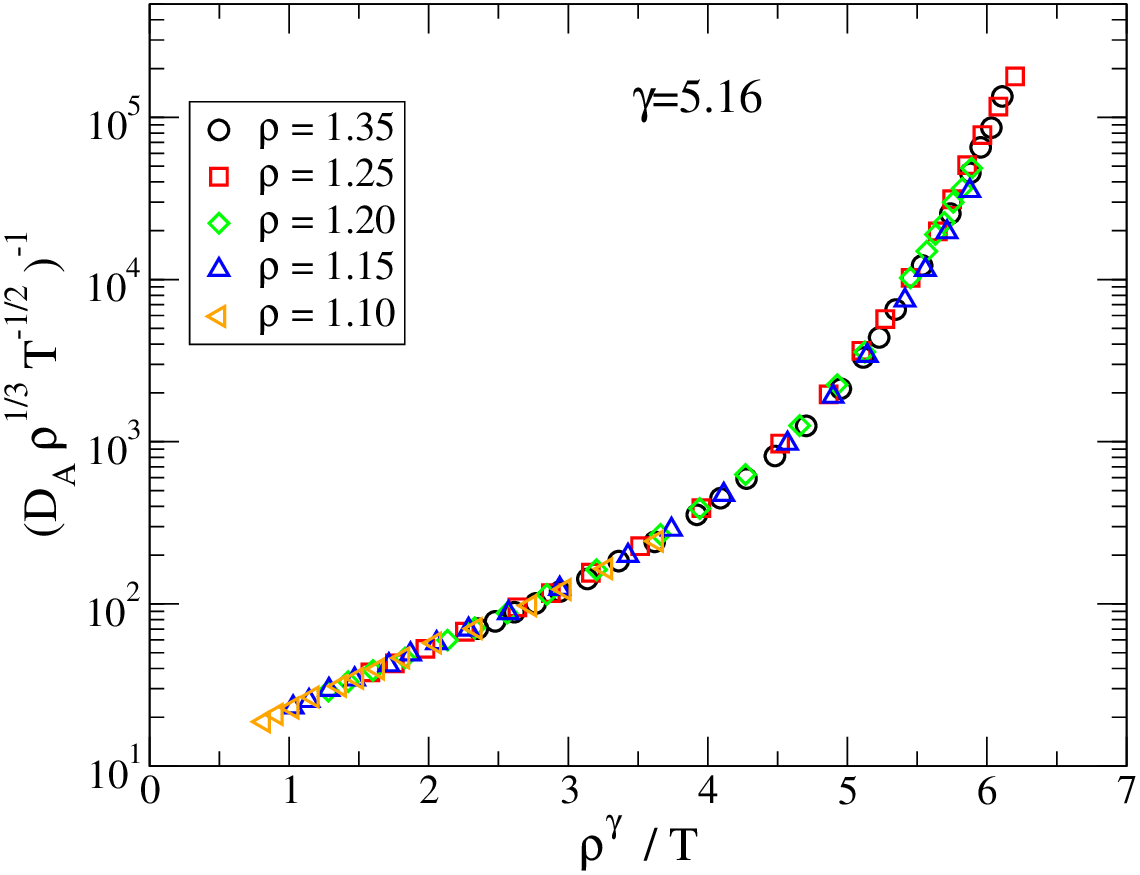} 
\hspace{3mm}
\includegraphics[scale=0.4]{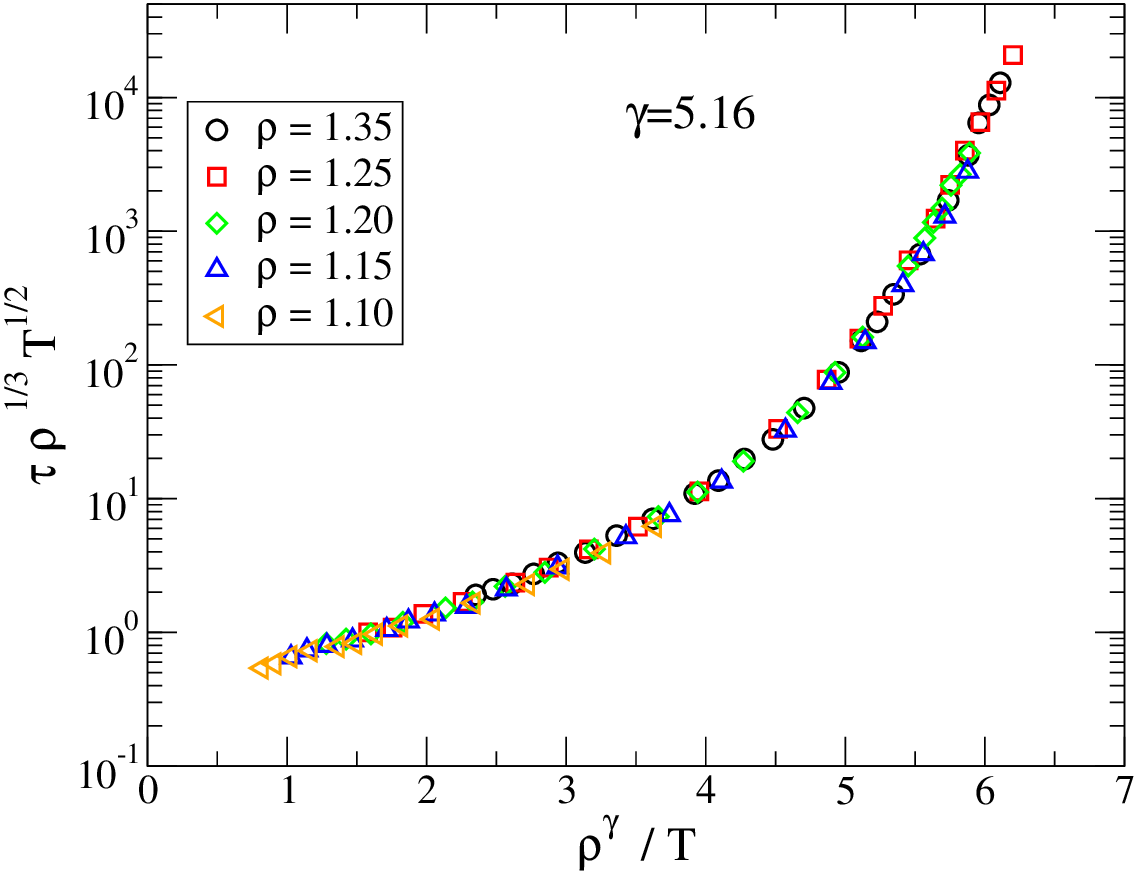}
\caption{Top row : The density temperature (DT) scaling in the KA model of the reduced diffusivity of $A$ particles $D_{A}^{*} = \rho^{1/3} (k_{B}T/m)^{-1/2} D_{A}$ and the reduced relaxation time $\tau^{*} = \rho^{1/3} (k_{B}T/m)^{1/2} \tau$. Bottom row : The quality of data collapse improves if state points with negative virials (Fig. \ref{fig:negative-virial}) are removed.}
\label{fig:DT-scaling}
\end{center}
\end{figure*}

\begin{figure}[h!]
\begin{center}
\includegraphics[scale=0.35]{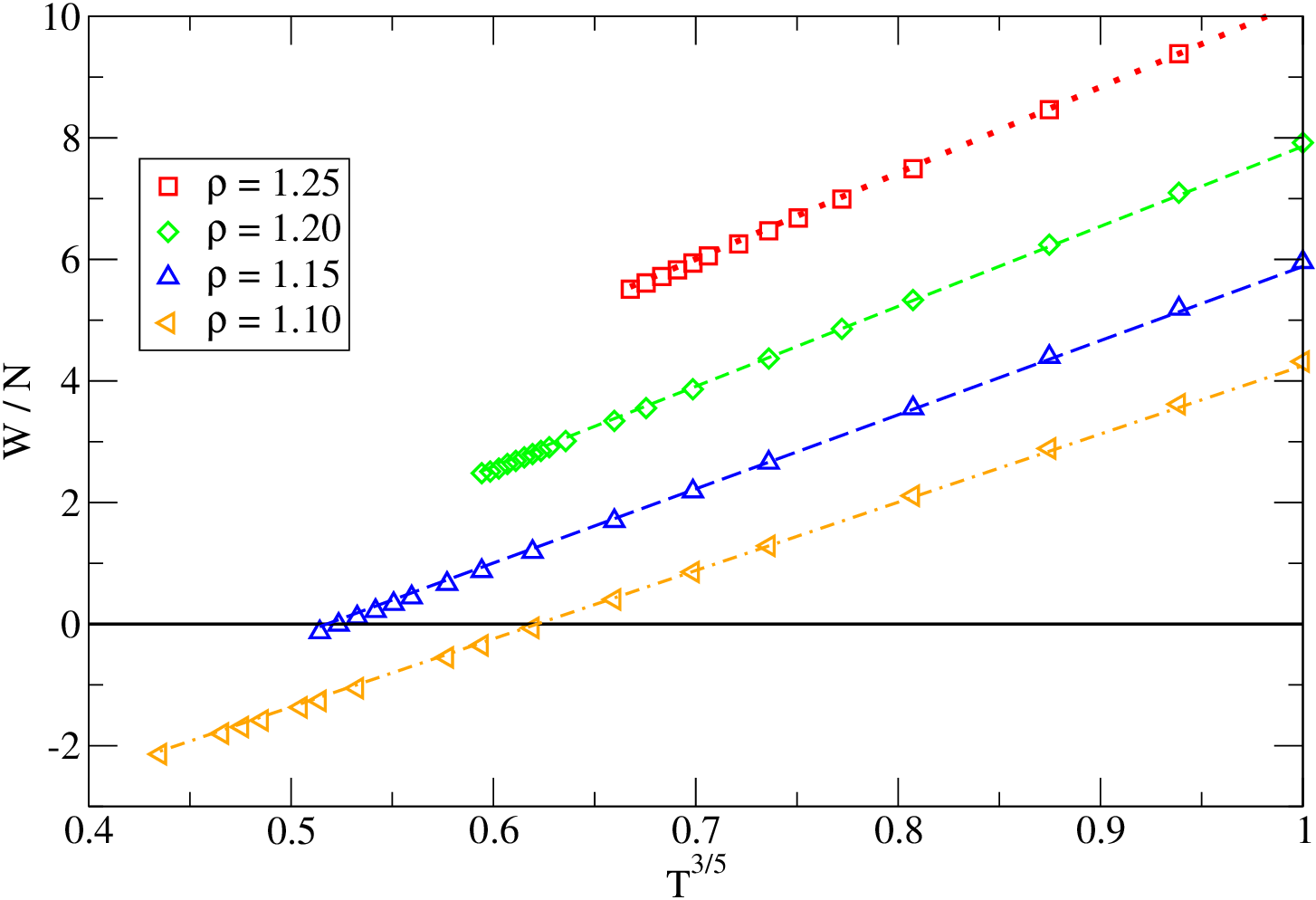}
\caption{Negative virial states occur at densities $\rho=1.10$ for temperatures $T<0.45$ and at $\rho=1.15$ for $T<0.35$. Lines are fits to the form $W/N = a + b T^{3/5}$ \cite{pap:scaling-5}. }
\label{fig:negative-virial}
\end{center}
\end{figure}

\section{Results: How strongly does the kinetic fragility depend on density?}





\begin{figure*}[t!]
\begin{center}
\includegraphics[scale=0.35]{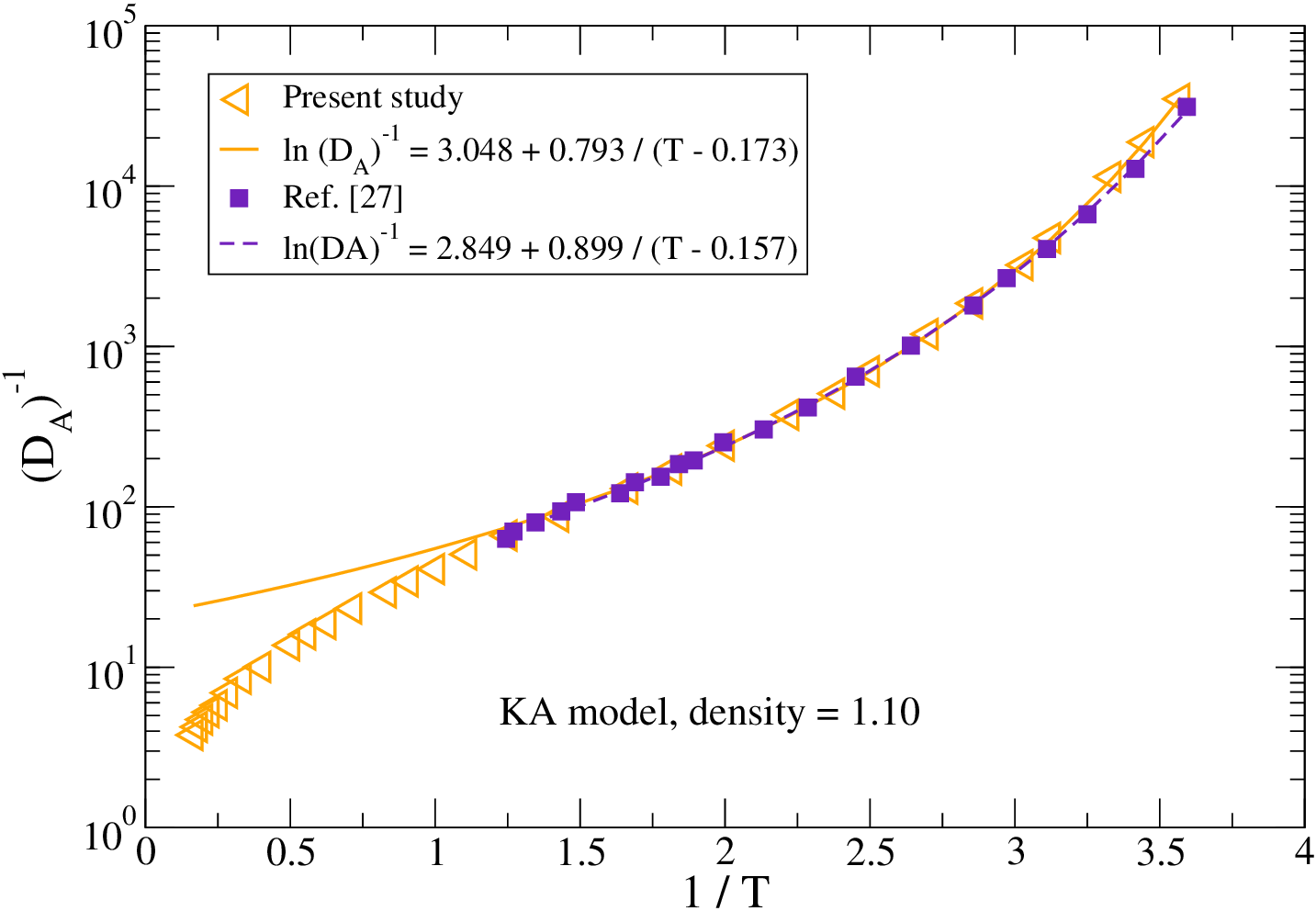}
\hspace{3mm}
\includegraphics[scale=0.35]{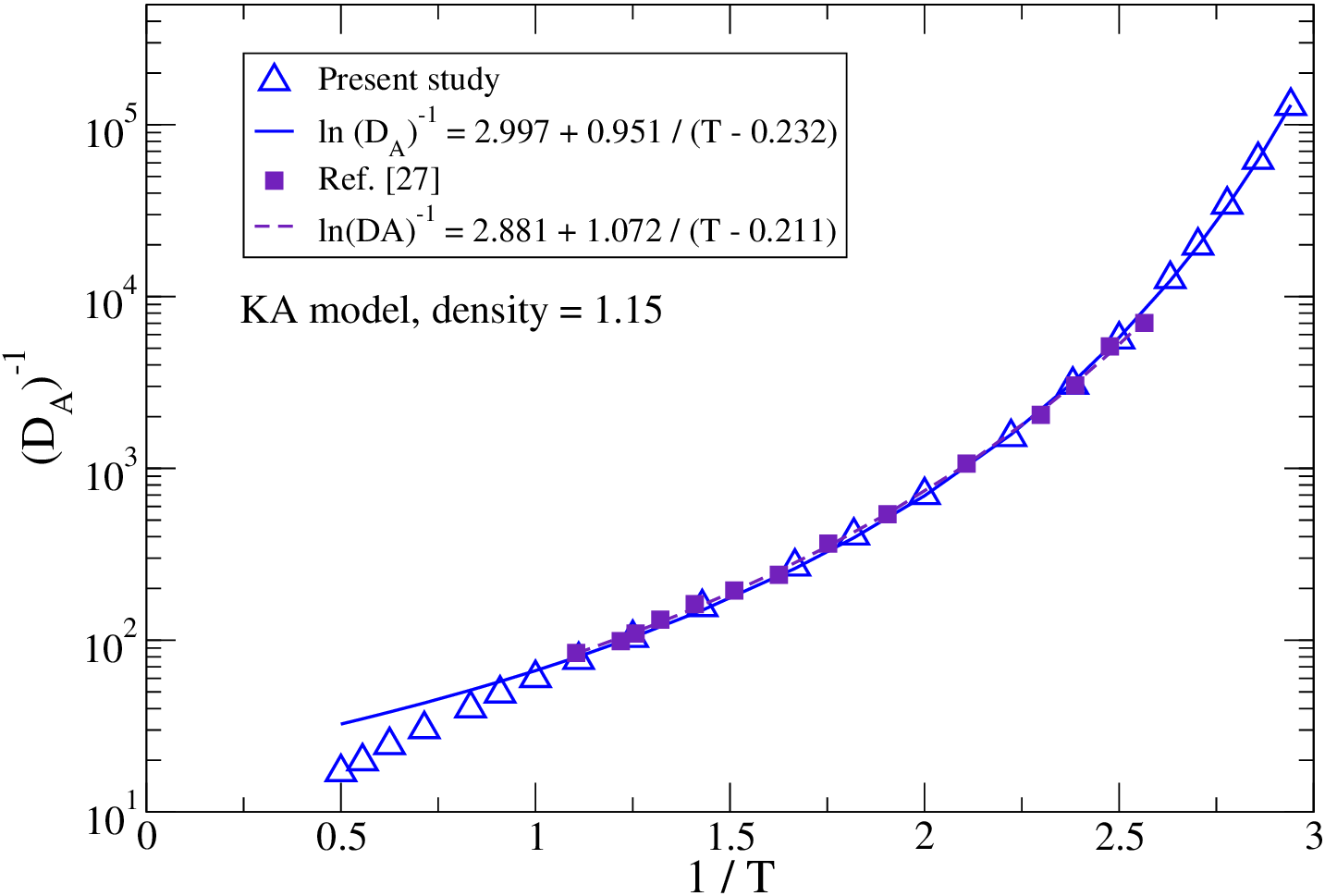}
\\\vspace{3mm}
\includegraphics[scale=0.35]{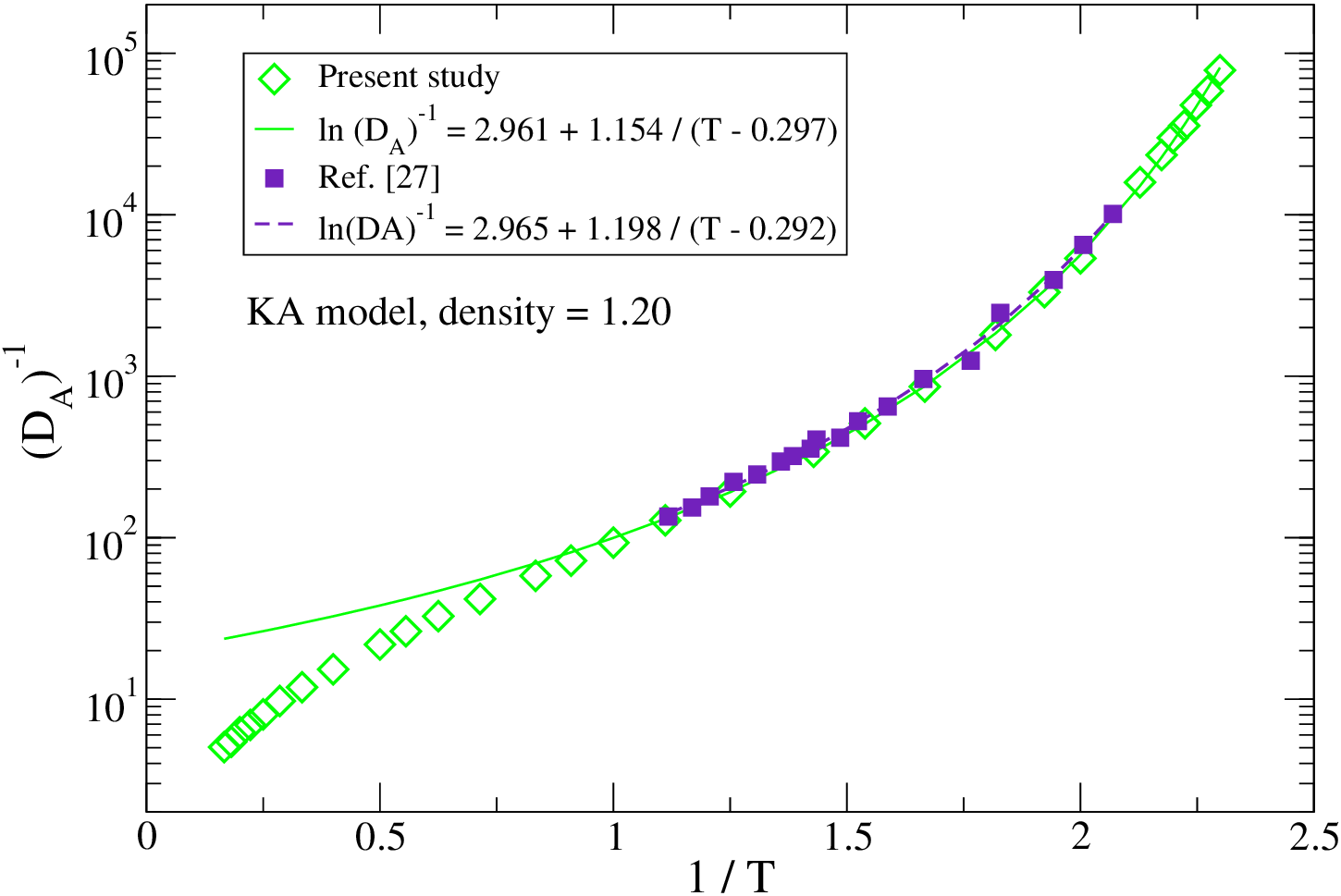}
\hspace{3mm}
\includegraphics[scale=0.35]{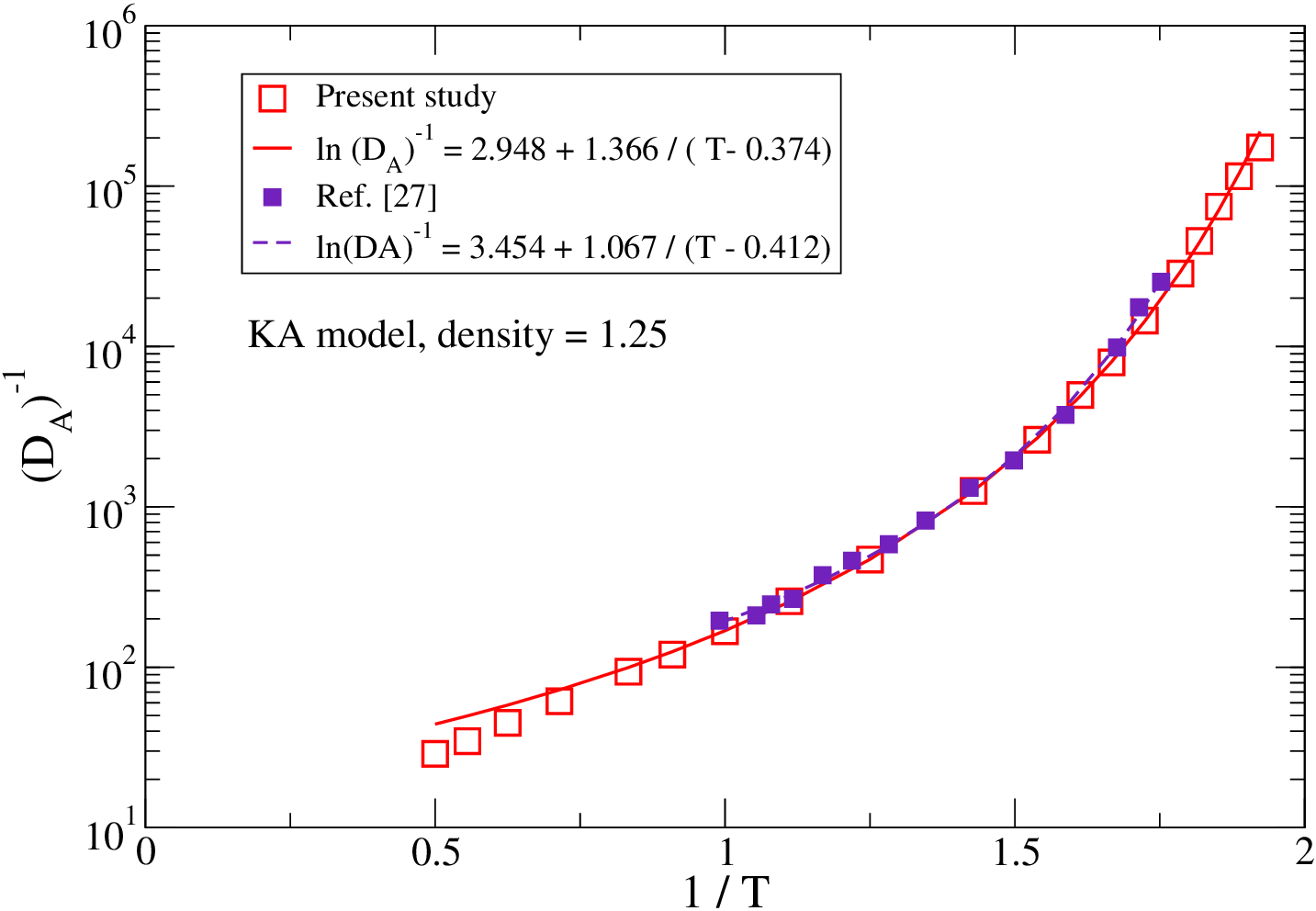}
\\\vspace{3mm}
\includegraphics[scale=0.35]{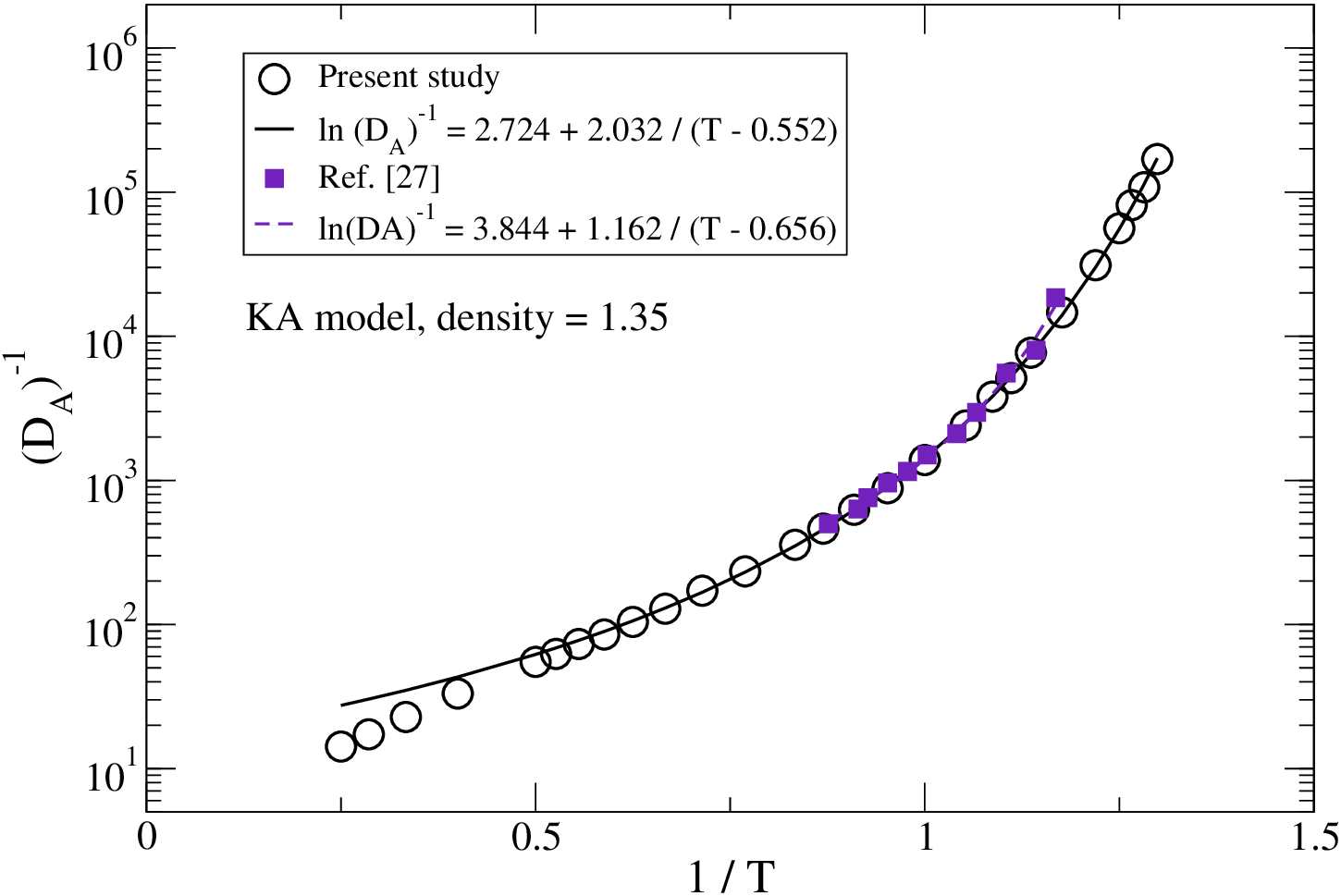}
\caption{Comparison of inverse of diffusivity of $A$-type particles ($D_{A}^{-1}$) computed for the KA model in the density range  $\rho= 1.10-1.35$ shows that the raw data for diffusivities obtained in the present study match well with the earlier work \cite{pap:Sc-Sastry}.}
\label{fig:3DKA-DA-compare}
\end{center}
\end{figure*}

\begin{figure*}[t!]
\begin{center}
\vspace{3mm}
\includegraphics[scale=0.35]{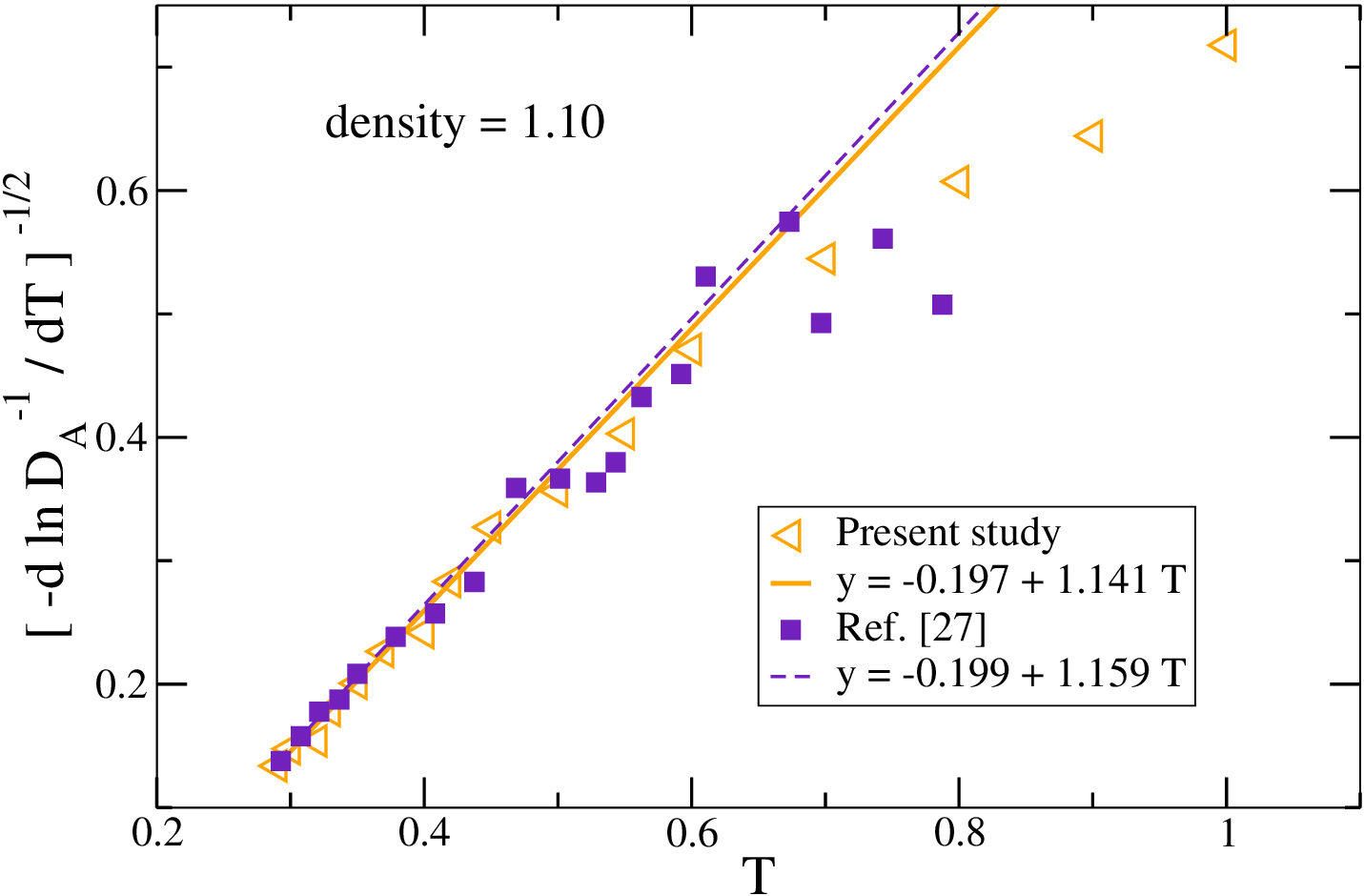}
\hspace{3mm}
\includegraphics[scale=0.35]{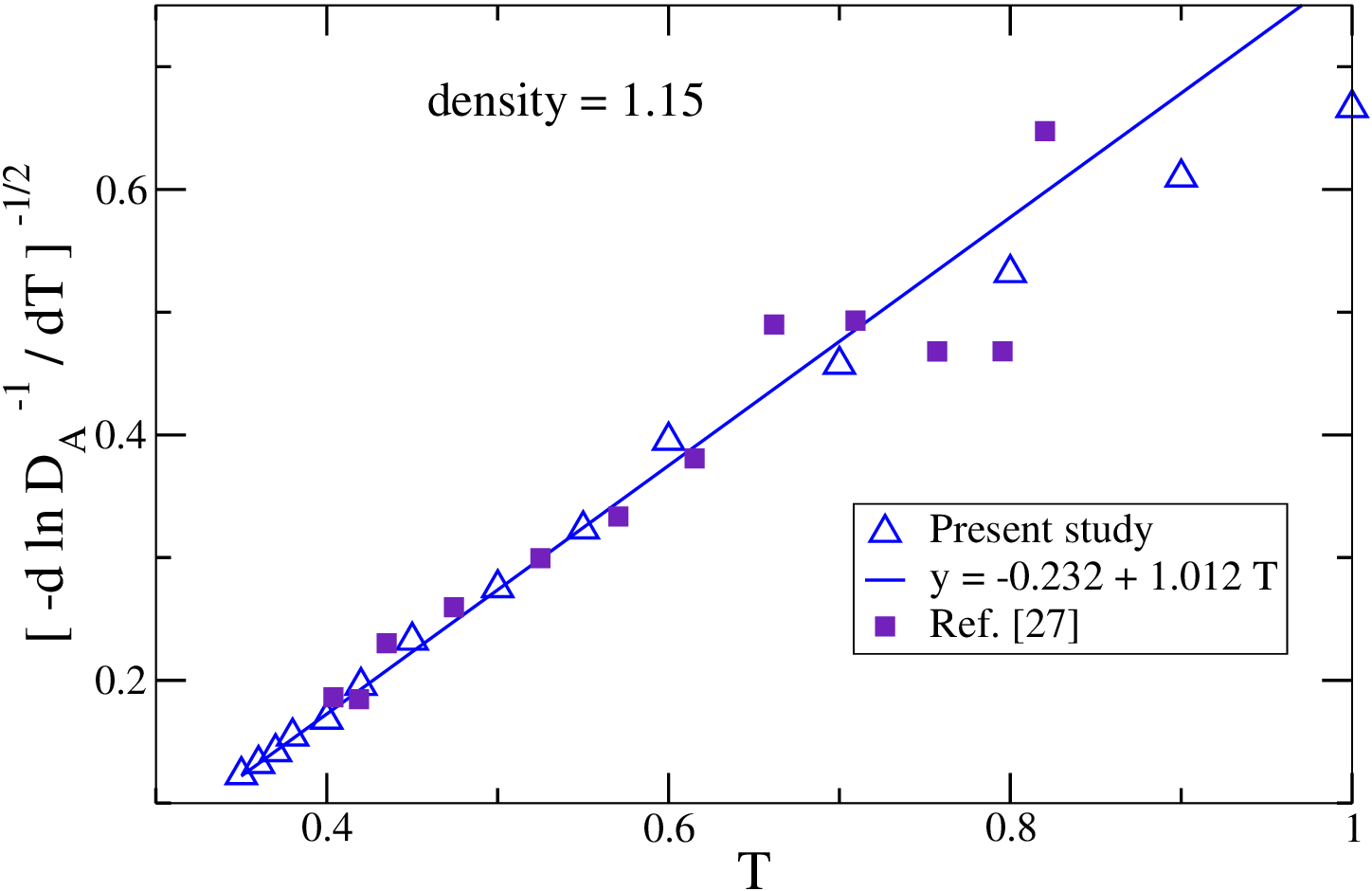}
\\\vspace{8mm}
\includegraphics[scale=0.35]{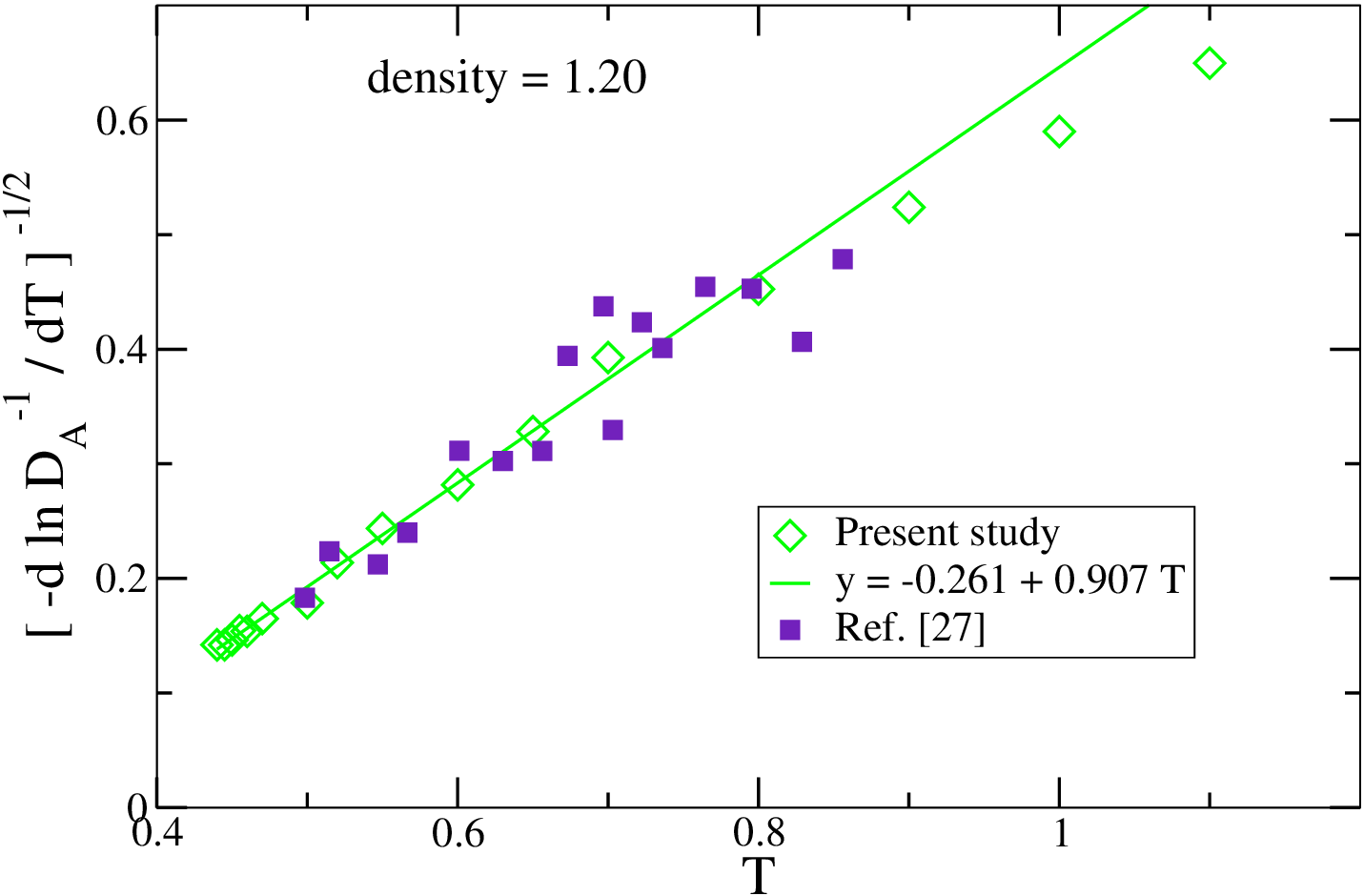}
\hspace{3mm}
\includegraphics[scale=0.35]{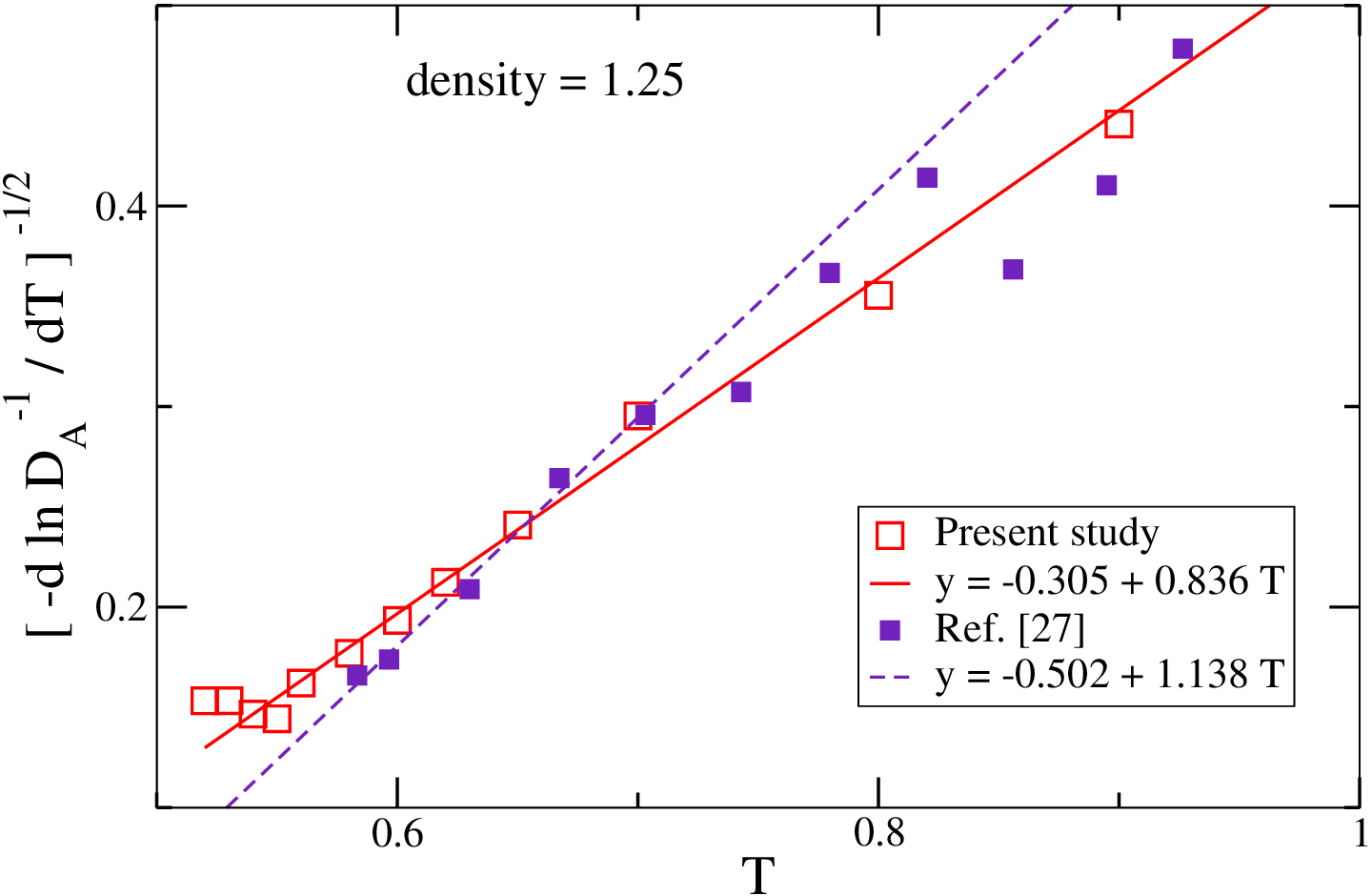}
\\\vspace{3mm}
\includegraphics[scale=0.35]{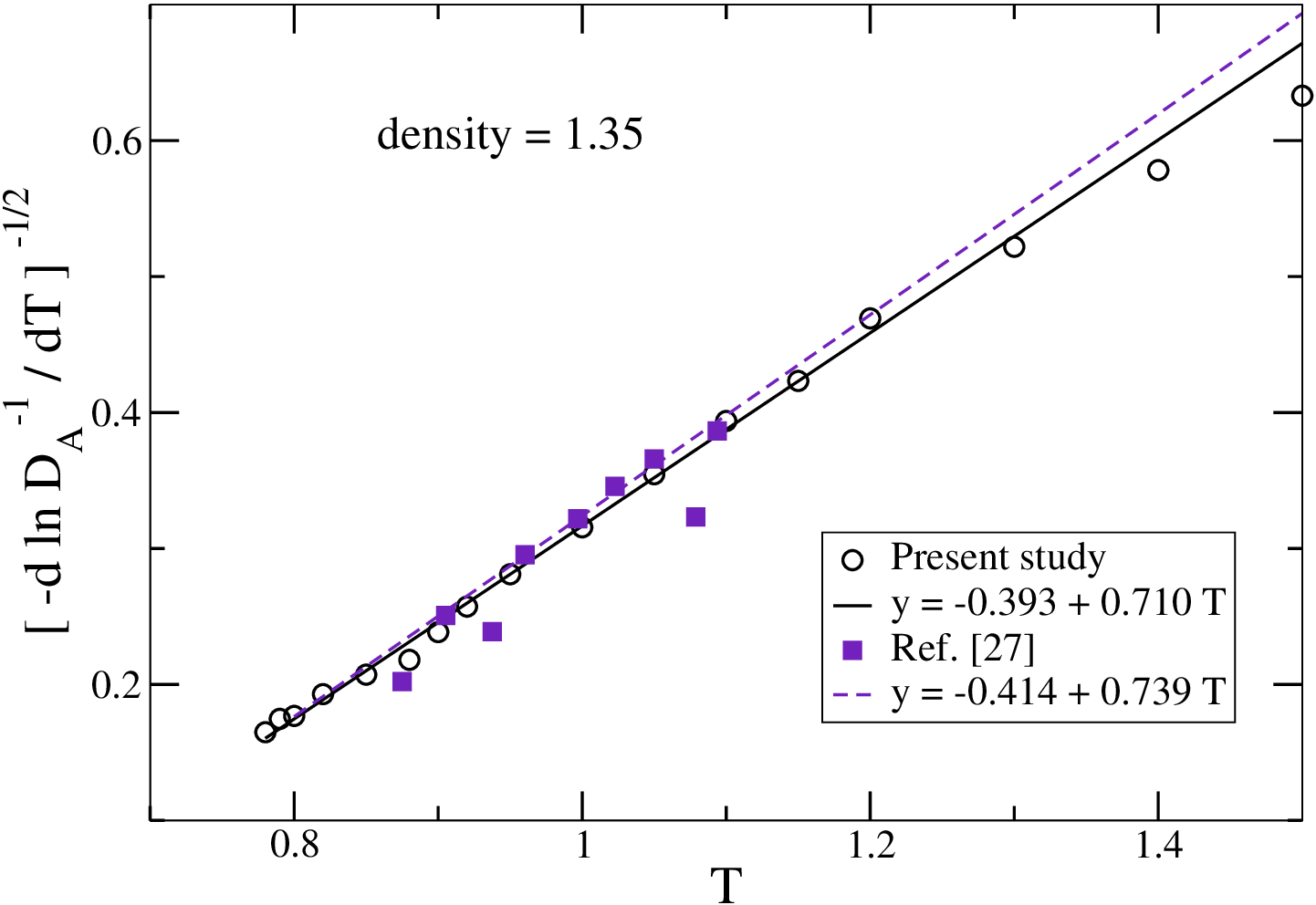}
\caption{Comparison of inverse diffusivity of $A$-type particles  in the present study with the earlier study \cite{pap:Sc-Sastry} in a representation suitable for Stickel analysis \cite{Stickelplot} (linearized VFT fits) in the density range $\rho = 1.10 - 1.35$ for the KA model. Lines are best fits of the form $\left[ - \frac{d \ln D_A}{dT} \right]^{-1/2} = \sqrt{\frac{K_{VFT}}{T_{VFT}}} T - \sqrt{K_{VFT}T_{VFT}}$ through low temperature data.}
\label{fig:3DKA-Stickelplot}
\end{center}
\end{figure*}

\subsection{Density-temperature scaling in the Kob-Andersen model}\label{sec:2}
For the KA model, the density-temperature scaling of relaxation times has been reported earlier \cite{pap:coslovich-roland-KA-DTscaling-tau}. Similar data collapse for diffusivities is also reported for another model with Lennard-Jones interaction \cite{pap:coslovich-roland-LJ-DTscaling-D}. We verify DT scaling for the KA model for \emph {both} relaxation times \emph{and} diffusivities in Fig. \ref{fig:DT-scaling}. Here the relaxation times are computed from the overlap function (see Section 3.1) and not from the intermediate scattering function. Note that the scaling exponent $\gamma=5.16$ is not a fit parameter but is \emph{predicted} from pressure-energy correlations \cite{KAIPL}. The data at low densities show deviation from the master curve. In Fig. \ref{fig:negative-virial} we show that at low densities, there are state points where the virial (W) is negative, hence the effective repulsive inverse power law potential may not be a good approximation to the KA model at those state points.  We show in Fig. \ref{fig:DT-scaling} that the quality of the data collapse improves if the negative virial states are excluded. 

From the above analysis, we conclude that the density temperature scaling in the KA model is valid to a very good approximation in the range of densities and temperatures studied here, with deviations at low density for low temperatures. Further, the deviation at low densities occur because of negative virial (pressure) states.

\subsection{Comparison of diffusivities and fragilities with earlier work}

\begin{figure}[h!]
\begin{center}
\includegraphics[scale=0.35]{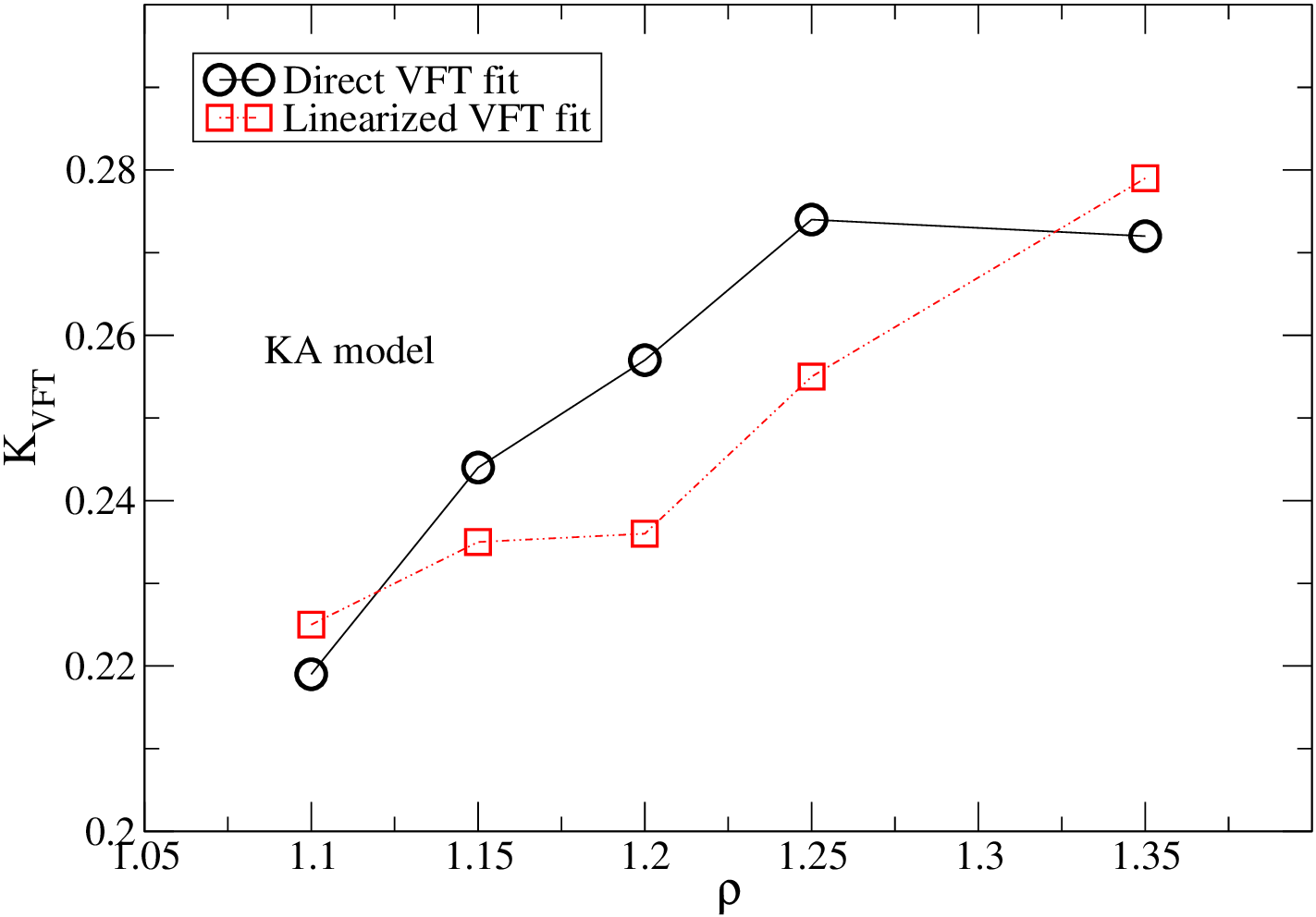}
\caption{Estimates of the kinetic fragility $K_{VFT}$ in the present study from direct VFT fits (Fig. \ref{fig:3DKA-DA-compare}) and from the Stickel plots (linearized VFT fits, Fig. \ref{fig:3DKA-Stickelplot}) are comparable to each other.}
\label{fig:compare-KVFT}
\end{center}
\end{figure}

\begin{figure}[h!]
\begin{center}
\includegraphics[scale=0.35]{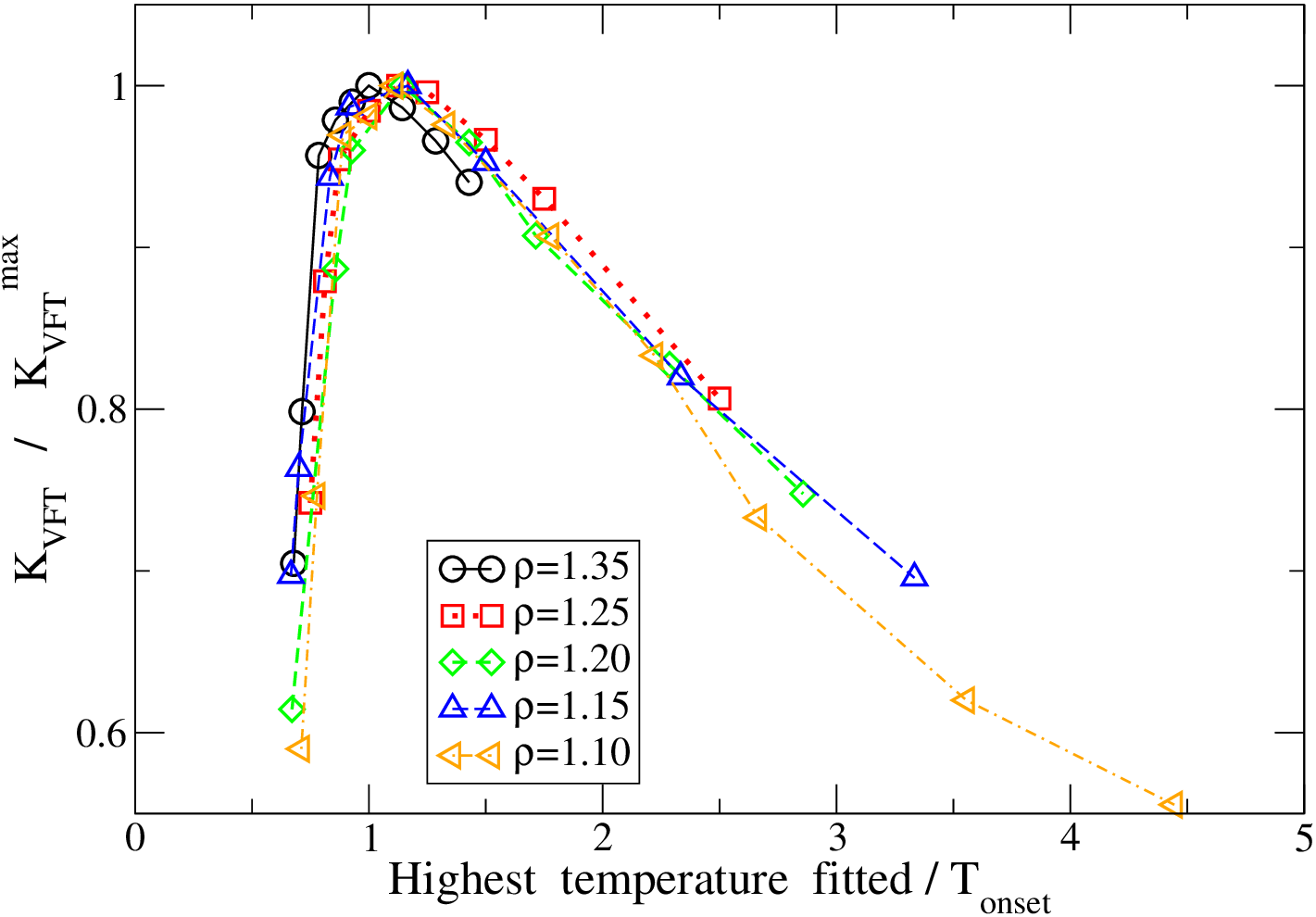}
\caption{Testing the sensitivity of kinetic fragility ($K_{VFT}$) estimates on the temperature range selected for VFT fitting. VFT fits to diffusivity data in the present study are done from the lowest available temperature to different choices of the highest temperature - which at each density vary from $\sim 70 \%$ of the onset temperature to the highest available temperature. The peak at each density occurs approximately at the onset temperature. Lines are guides to eyes. }
\label{fig:KVT-vs-Trange-new}
\end{center}
\end{figure}

If the density-temperature (DT) scaling is exact, then the kinetic fragility should be independent of density while the thermodynamic fragility should show power law dependence on density. The latter follows from the prediction of the thermodynamic theory of DT scaling that the configuration entropy $S_c$ is a function of $\rho^\gamma / T$. Hence from Eqn. \ref{eqn:def-KAG}, $S_c = \frac{K_T}{T}( T / T_K - 1) = S_c (\rho^\gamma / T) \Rightarrow K_T \sim \rho^\gamma$.  Given that DT scaling holds to a very good approximation in the KA model, we expect that the density dependence of the kinetic fragility should be much weaker than that of the thermodynamic fragility. However, in Ref.  \cite{pap:AG-Sastry} it was found that in the density range ($\rho=1.10-1.35$) studied here for the KA model, the kinetic fragility  changed approximately by a factor of 3 and the thermodynamic fragility approximately by a factor of 3.5 {\it i.e.} the density dependences were comparable.  Hence, to resolve this apparent contradiction, we re-analyze the kinetic and the thermodynamic fragility for the 3D KA model in the present study.  

First, we compare the diffusivity data in the present study with Ref. \cite{pap:Sc-Sastry} in Fig. \ref{fig:3DKA-DA-compare} and have found that they agree well. 

Next, we estimate kinetic fragility values from the diffusivity data using Eqn. \ref{eqn:kinfr2} by performing VFT fits : $\ln D_{A}^{-1} = \ln ( D_A^{-1} )_0 +  \frac{1}{ K_{VFT}(T / T_{VFT} - 1)}$.  The fitting curves are shown in Fig. \ref{fig:3DKA-DA-compare}. We see that VFT fits are of comparable quality for both the present study and Ref.  \cite{pap:Sc-Sastry}. Hence it is not evident that the kinetic fragility estimate in the present study is more (or less) reliable than Refs.  \cite{pap:Sc-Sastry,pap:AG-Sastry}. 

Hence we analyze the diffusivity data more critically following the method of  Stickel \textit{et al.} \cite{Stickelplot}.  In this method, the temperature dependence of diffusivity is represented in  a way which reveals any noise in data more easily than the direct VFT fit. The basic strategy is to is to \emph{linearize} the VFT formula with respect to $T$ : $\left[ - \frac{d \ln D_A}{dT} \right]^{-1/2} = \sqrt{\frac{K_{VFT}}{T_{VFT}}} T - \sqrt{K_{VFT}T_{VFT}}$. Computing the derivative of diffusivity with respect to $T$ makes the linearized form particularly sensitive to the noise present in the diffusivity data. We also note that the Arrhenius behaviour in this representation would correspond to a straight line passing through the origin. We show the $T$ dependence of  $\left[ - \frac{d \ln D_A}{dT} \right]^{-1/2}$ obtained in the present study in Fig. \ref{fig:3DKA-Stickelplot}.  For comparison, we also show data from Ref.  \cite{pap:Sc-Sastry}. For both sets, the temperature derivative is computed numerically using the central difference formula. We see from Fig. \ref{fig:3DKA-Stickelplot} that both data sets show linear behaviour only at low temperatures. However, in comparison to Ref.  \cite{pap:Sc-Sastry}, data in the present study is less noisy and remains linear for a bigger range of temperature. Hence we conclude that estimates of kinetic fragility obtained in the present study is more reliable than in Refs. \cite{pap:Sc-Sastry,pap:AG-Sastry}.

The estimates of the kinetic fragility obtained in the present study from both the direct VFT fit and from the Stickel analysis are shown in Fig. \ref{fig:compare-KVFT}.  We see the values from the two methods are mutually comparable. Further, in the density range $\rho = 1.10 - 1.35$, the kinetic fragility changes by a factor of roughly $1.3$.  In other words, the variation of the kinetic fragility with density in the present study is significantly \emph{less} compared to Refs. \cite{pap:Sc-Sastry,pap:AG-Sastry}. 

The main differences between the present study with Refs. \cite{pap:AG-Sastry} are that (i) the system size  is bigger in the present study (N=1000) compared to Ref.  \cite{pap:AG-Sastry} (N=256 ), (ii) the temperature range is also wider in the present study and (iii) the runlengths are longer typically by factors of 2-6 at low temperatures and by a factor of 20 at high temperatures. To understand the difference between the present study and Ref.  \cite{pap:AG-Sastry} in the $K_{VFT}$ values despite good agreement in the $D_{A}^{-1}$ data, we test the sensitivity of the estimate of the kinetic fragility from the direct VFT fit on the choice of temperature range for fitting in Fig. \ref{fig:KVT-vs-Trange-new}.  We fit data from the lowest available temperature to different choices of the highest temperature. The choice of the highest temperature at each density varies from $\sim 70 \%$ of the onset temperature (obtained from the T-dependence of the average inherent structure energy) to the highest available temperature. Fig. \ref{fig:KVT-vs-Trange-new} clearly shows that the estimate of kinetic fragility from VFT fit is sensitive to the choice of the temperature range for fitting.  Based on the observations from Figs. \ref{fig:3DKA-Stickelplot} and \ref{fig:KVT-vs-Trange-new}, we attribute the relatively high values of kinetic fragility obtained in the earlier study to two factors, namely (a) less reliable estimate of diffusivity at low temperatures and (b) relatively shorter range of temperature studied.

\begin{figure}[h!]
\begin{center}
\includegraphics[scale=0.35]{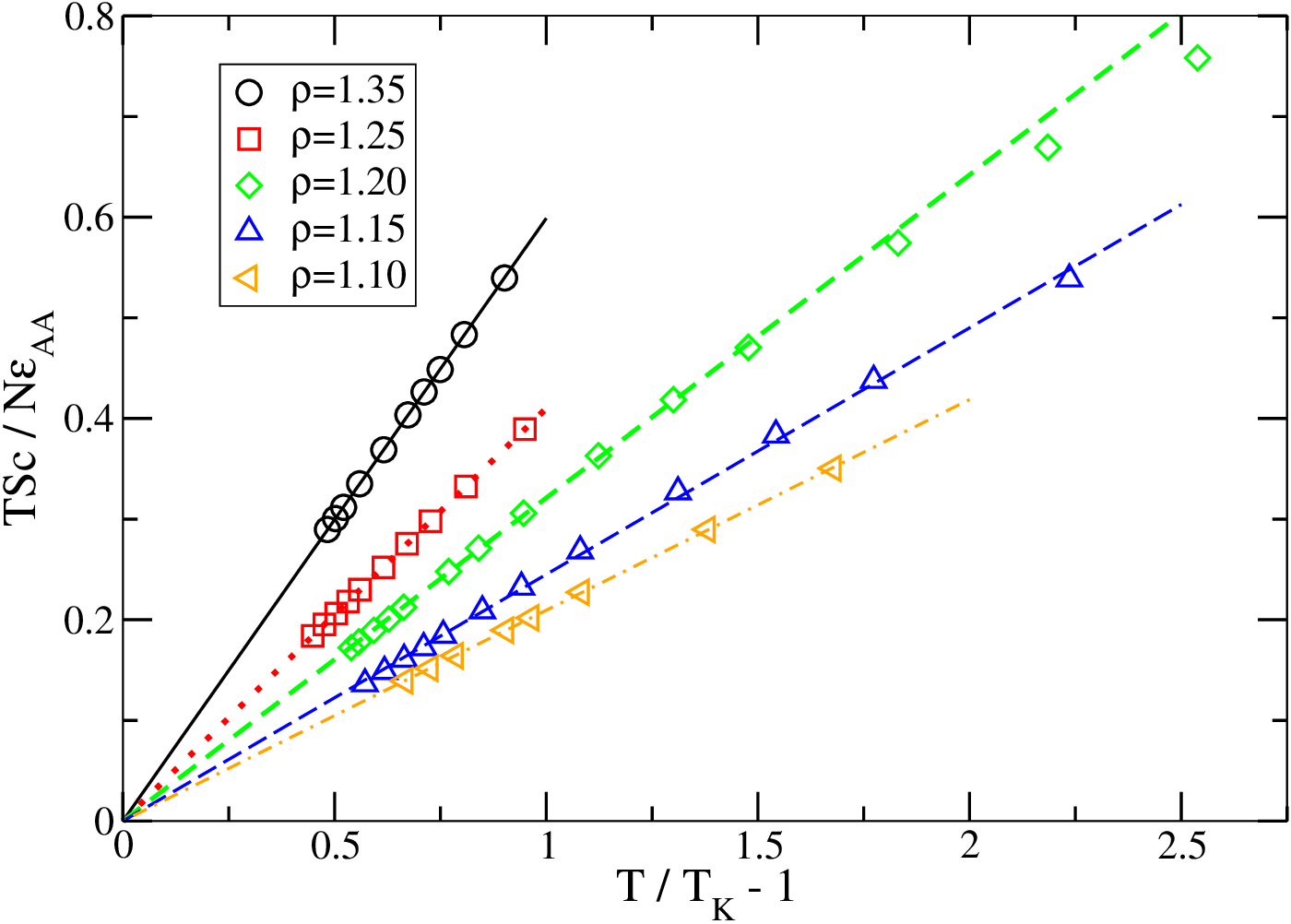}
\caption{The temperature dependence of $TS_{c}$ obtained in the present study at different densities for the KA model. The thermodynamic fragility $K_T$ at different densities are obtained from this plot using Eqn. \ref{eqn:def-KAG}.}
\label{fig:TSc-T-diff-density}
\end{center}
\end{figure}

\begin{figure}[h!]
\begin{center}
\includegraphics[scale=0.33]{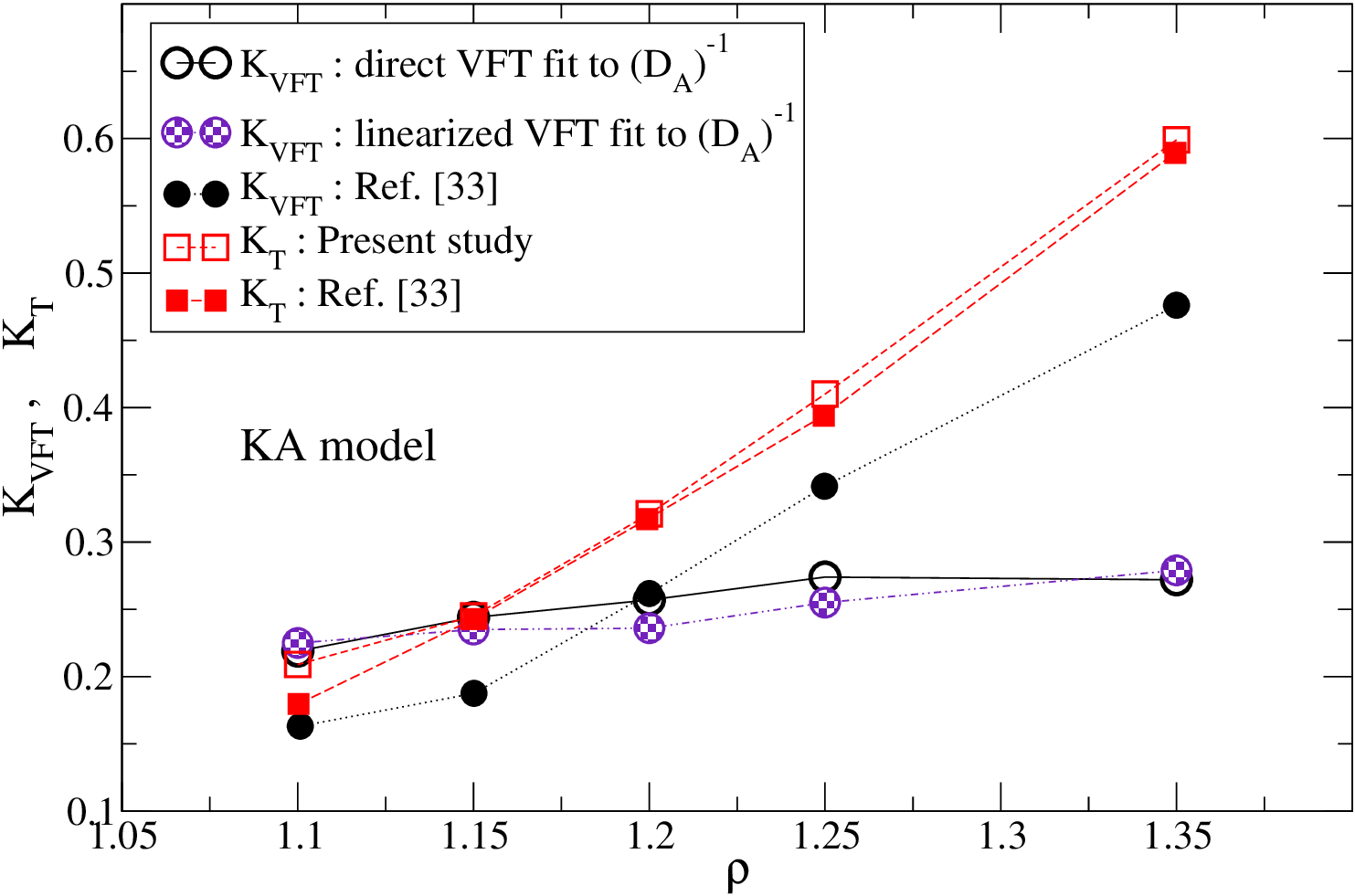}
\caption{The density dependence of $K_{VFT}$ is much weaker in the present study compared to Ref.  \cite{pap:AG-Sastry}. The density dependence of the thermodynamic fragility $K_{T}$ is comparable to the earlier study.}
\label{fig:KVT-KT-vs-density}
\end{center}
\end{figure}

The thermodynamic fragility values in the present study are estimated using Eqn. \ref{eqn:def-KAG} from the temperature dependence of $TS_{c}$ and are shown in Fig.  \ref{fig:TSc-T-diff-density}. Finally, Fig. \ref{fig:KVT-KT-vs-density} summarizes the comparison of the density dependence of the kinetic and the thermodynamic fragilities for the KA model in the present study and Ref. \cite{pap:AG-Sastry} and shows that (i) the density dependence of the kinetic fragility is much weaker in the present study and (ii) the density dependence of the thermodynamic fragility in the present study agrees well with Ref.  \cite{pap:AG-Sastry}.

\begin{figure}[h!]
\begin{center}
\includegraphics[scale=0.35]{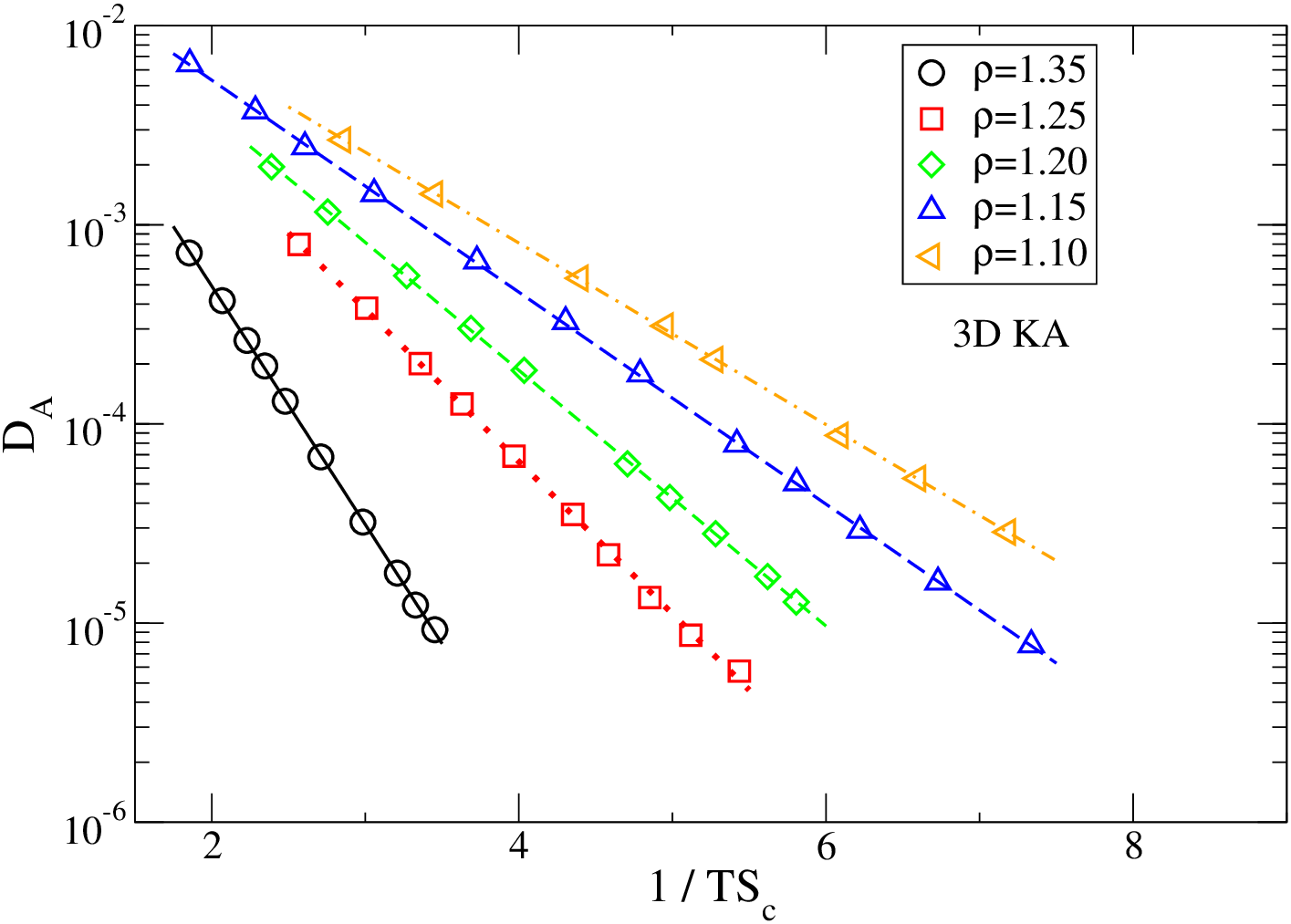}
\\\vspace{5mm}
\includegraphics[scale=0.35]{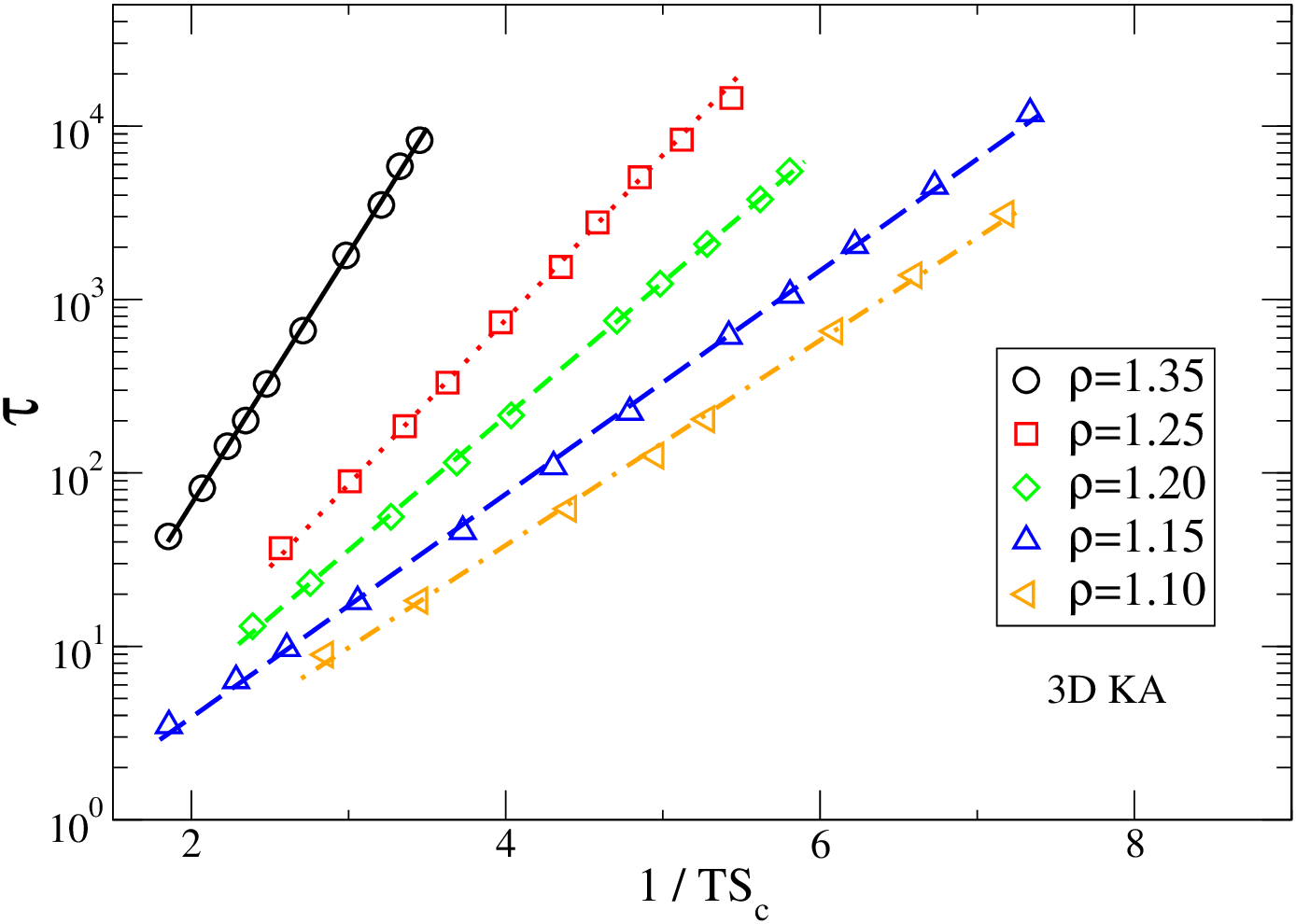}
\caption{The Adam Gibbs plots for the 3D KA model in terms of diffusivities and relaxation times at different densities obtained in the present study.}
\label{fig:AG-tauqt-DA}
\end{center}
\end{figure}

\subsection{Density dependence of fragilities explained from DT scaling}
After establishing that the density dependence of the kinetic fragility is rather weak for the KA model, we try to rationalize the density dependence from DT scaling. First we verify in Fig. \ref{fig:AG-tauqt-DA} using both diffusivity and relaxation time that the Adam Gibbs (AG) relation (see Eqn. \ref{eqn:AG}) is valid at all the densities studied here so that the kinetic and the thermodynamic fragility can be related to each other by Eqn. \ref{eqn:KT-KVFT}. The values of the AG coefficient $A_{\tau}$ (from relaxation time) and $A_{D}$ (from diffusivity) are estimated from Fig. \ref{fig:AG-tauqt-DA}. Now according to the prediction of the thermodynamic theory for DT scaling \cite{pap:scaling-4}, 
both relaxation time and configurational entropy show DT scaling : $\tau = f(\rho^\gamma / T)$ and $S_c = g(\rho^\gamma / T)$. Hence from the AG relation : $\tau = \tau_0 \exp \left(A_{\tau} / TS_c \right)$, we expect a power law density dependence for the AG coefficient : $A_{\tau} \sim \rho^{\gamma_{A, \tau}}, A_{D} \sim \rho^{\gamma_{A,D}}$. Similarly as discussed before, from the $T$ dependence of $S_c = = \frac{K_T}{T}( T / T_K - 1) = S_c (\rho^\gamma / T)$ we expect $K_T \sim \rho^{\gamma_{K_T}}$.  Here we use different suffix for scaling exponent $\gamma$ obtained from different quantities to emphasize that the estimates can be different in general. However, from Eqn. \ref{eqn:KT-KVFT}, we obtain an estimate of the kinetic fragility as $K_{AG} (\rho) = K_T (\rho)/ A (\rho)$. If $K_{AG}$ is a good estimate for the measured kinetic fragility $K_{VFT}$ then we expect $\gamma_{K_T} \sim \gamma_{A}$ \textit{i.e.} the density dependence of $K_T$ and $A$ approximately cancel each other. 

\begin{figure}[t!]
\begin{center}
\includegraphics[scale=0.33]{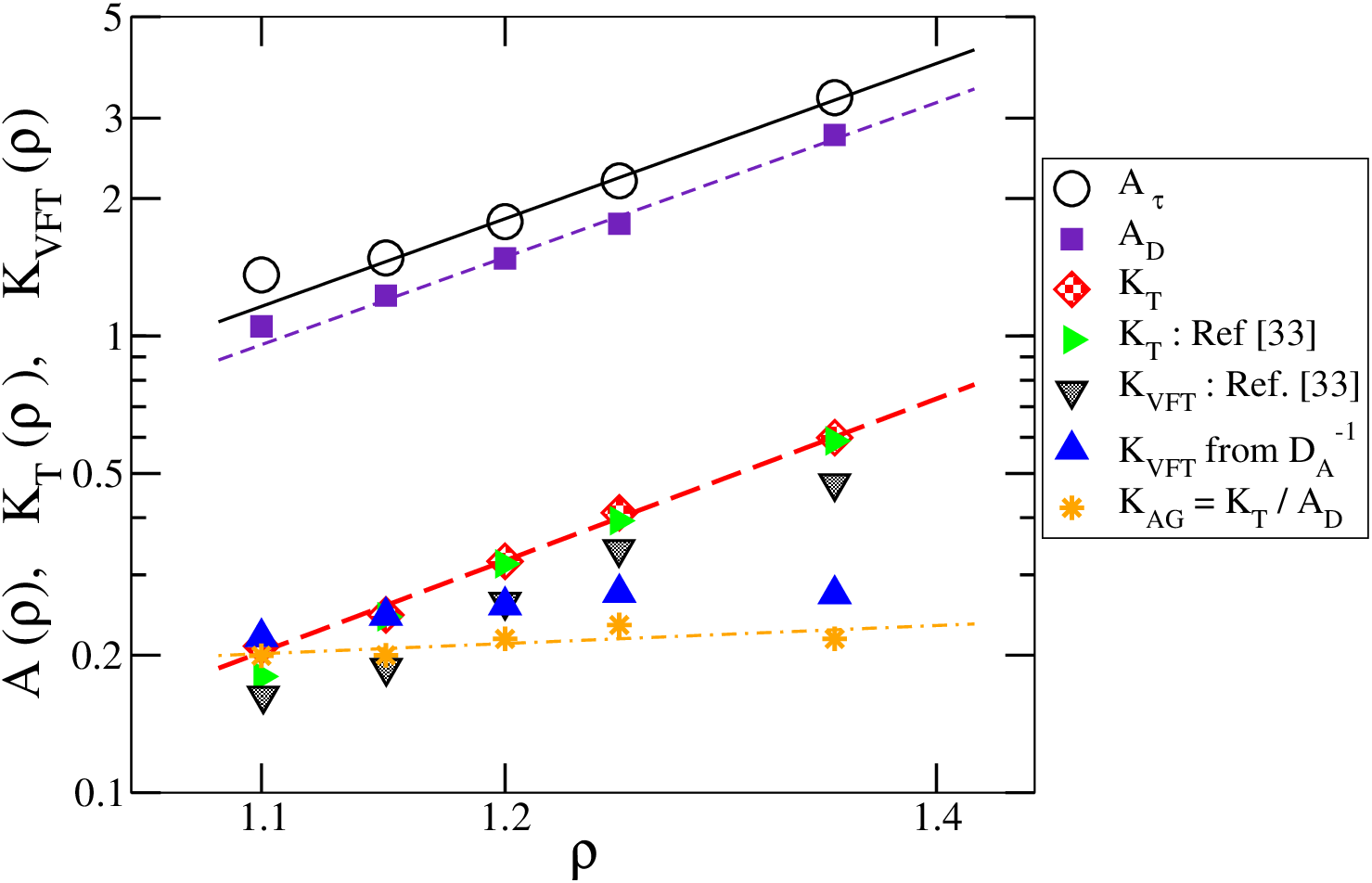}
\\\vspace{3mm}
\includegraphics[scale=0.33]{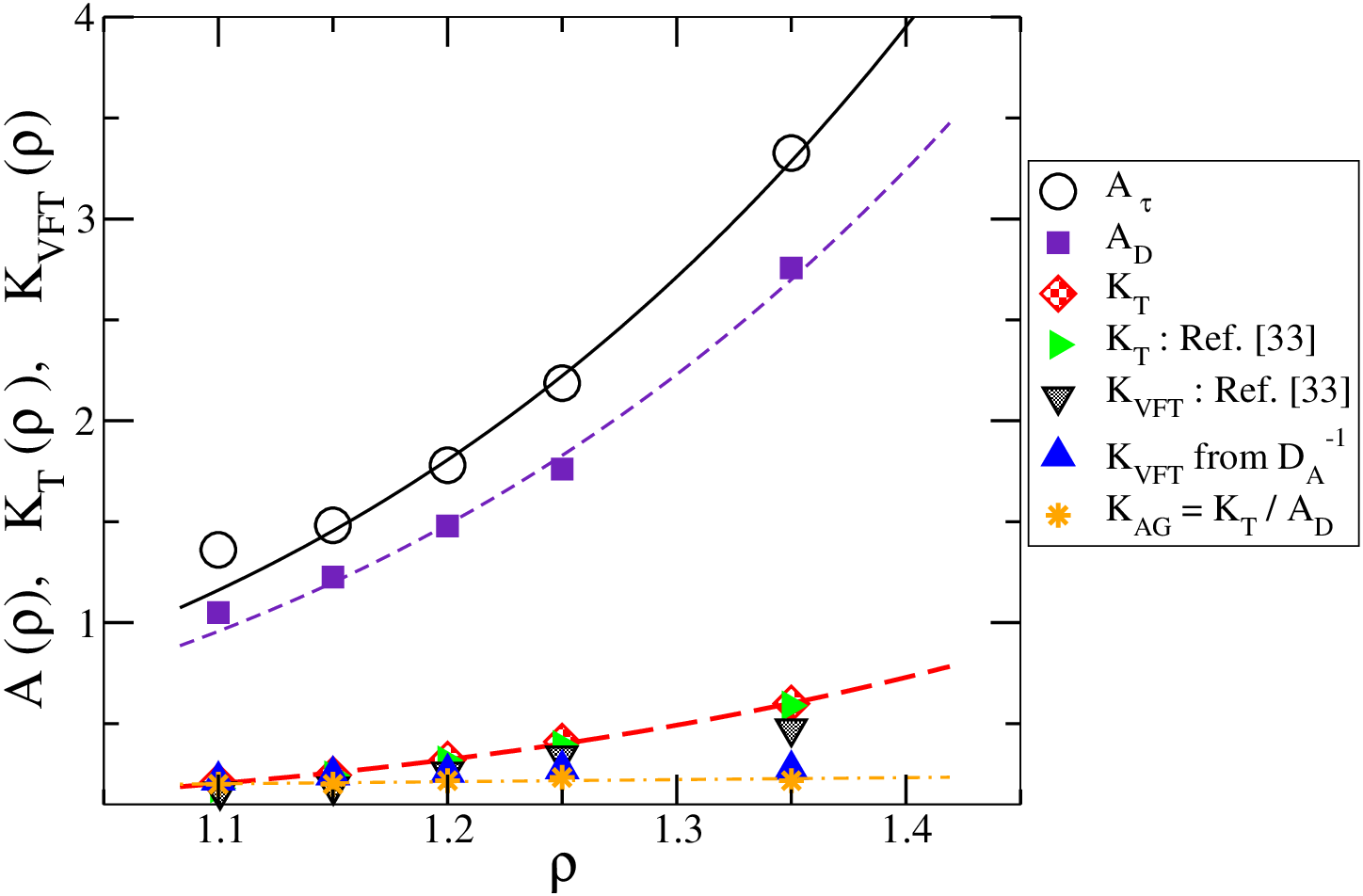}
\caption{Comparison in log-log (\emph{upper panel}) and in linear (\emph{lower panel}) scale of the density dependence of (i) the Adam Gibbs coefficient $A_{\tau}$ (from relaxation time data) and $A_{D}$ (from diffusivity data) (both sets shifted by constant factors), (ii) the thermodynamic fragility $K_T$, (iii) the kinetic fragility $K_{VFT}$ obtained from diffusivity data (shifted by constant factor) and (iv) the ratio $K_{AG} = K_{T} / A_{D}$ for the diffusivity data (shifted by constant factor). Also shown for comparison are the fragility values from Ref.  \cite{pap:AG-Sastry}. The scaling exponents $\gamma_{A,\tau} = 5.08 $ and $\gamma_{A, D} = 5.06$ are obtained from the density ($\rho$) dependences of the AG coefficients $A_{\tau}$ and $A_{D}$ respectively. The fit line through $K_{AG}$ has a scaling exponent $\gamma_{K_{AG}} = 0.59$ indicating that the density dependence of $K_{AG}$ is weak. }
\label{fig:compare-frag-present-study}
\end{center}
\end{figure}

The density dependence of (i) the Adam Gibbs coefficient $A$, (ii) the thermodynamic fragility $K_{T}$,  (iii) the kinetic fragility $K_{VFT}$ and (iv) the estimated kinetic fragility $K_{AG} = K_T / A$ for the 3D KA model obtained in the present study are summarized in Fig. \ref{fig:compare-frag-present-study}. We see that (i) the power law density dependence of $A, K_T$ describes data well except at the lowest density and (ii) The density dependence of the estimate $K_{AG}$ is weak and roughly proportional to the measured kinetic fragility  $K_{VFT}$. 

\begin{figure}[h!]
\begin{center}
\includegraphics[scale=0.33]{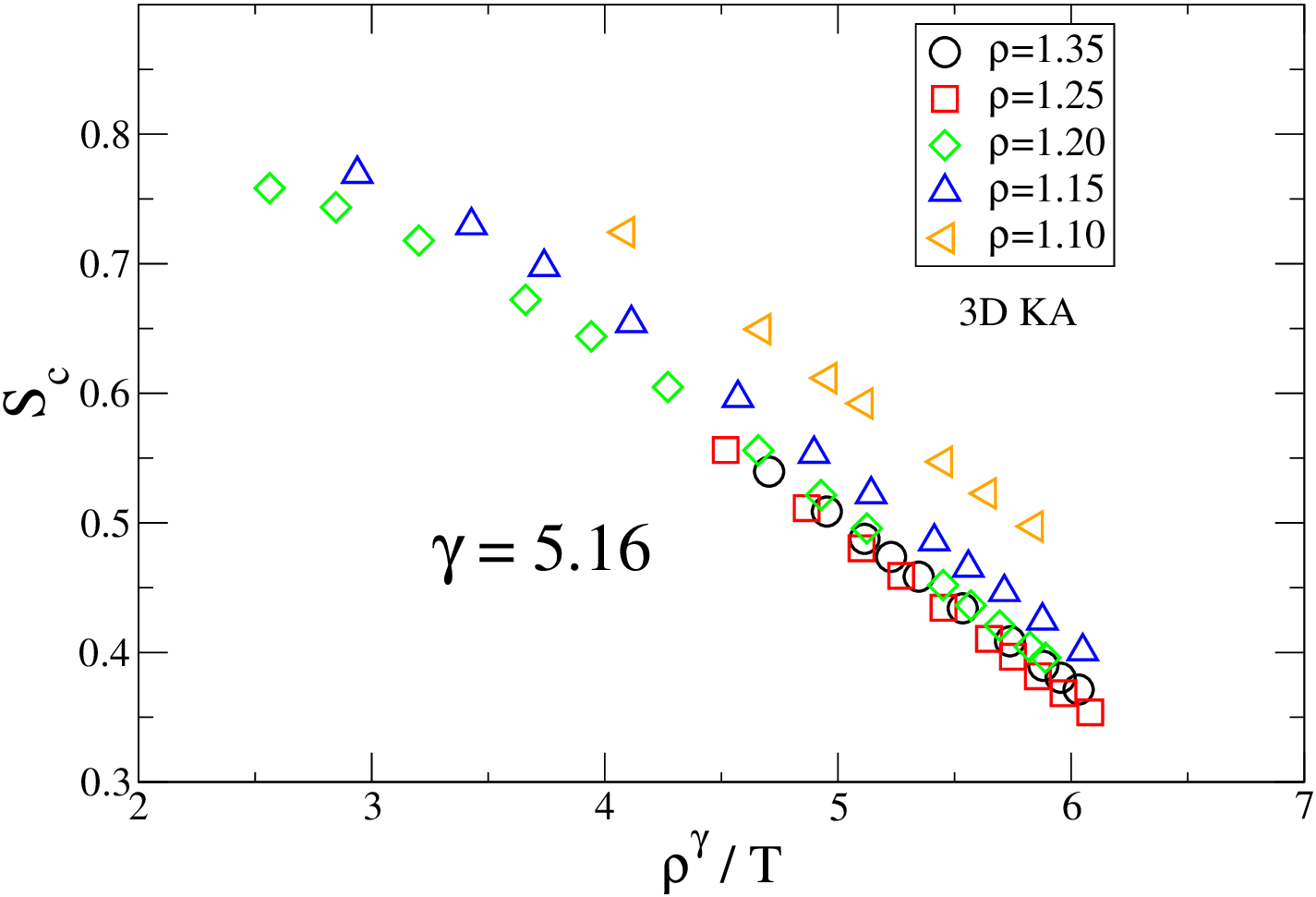}
\\\vspace{5mm}
\includegraphics[scale=0.33]{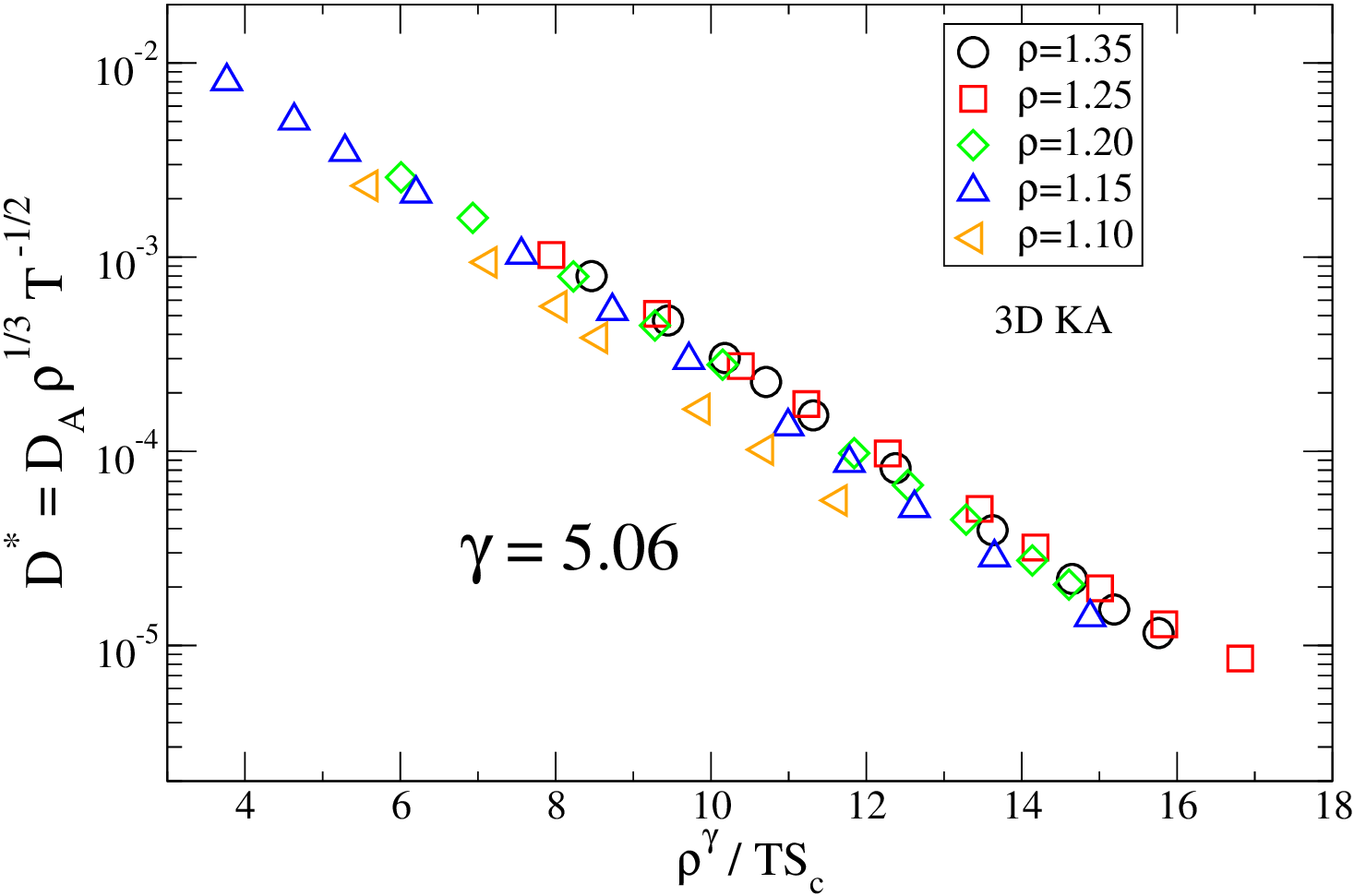}
\caption{\emph{Upper panel:} Testing if $S_{c}$ is an invariant under the density - temperature scaling with the scaling exponent $\gamma=5.16$ predicted from the pressure energy correlation. We find reasonably good data collapse at high densities. \emph{Lower panel : } Density - temperature scaling of the Adam Gibbs relation between the diffusivity and the configurational entropy.  Note that all data points at the lowest density $\rho =1.10$ in both figures are negative virial states. }
\label{fig:scaled-AG-plot}
\end{center}
\end{figure}

Next, we verify in Fig. \ref{fig:scaled-AG-plot} the prediction \cite{pap:scaling-4} that the configurational entropy $S_c$ is a function of the form  $S_{c}(\rho^{\gamma} / T)$  using the scaling exponent $\gamma = 5.16$ predicted from pressure-energy correlations \cite{KAIPL}. We see that except at the lowest density, one gets a reasonably good data collapse. We point out that all the data points at the lowest density $\rho = 1.10$ correspond to negative virial states. This is a possible reason for the observed deviation at the lowest density.  Finally, in Fig. \ref{fig:scaled-AG-plot} we also show that the Adam Gibbs plots at different densities can be collapsed on a master curve which is expected from DT scaling. Here we use the scaling exponent $\gamma = 5.06$ which is obtained from the density dependence of the AG coefficient $A_D$ estimated from diffusivity.

\section{Conclusions}
In the present study  on a model glass-former, the Kob-Andersen (KA) model, we have verified that to a very good approximation both diffusivity and relaxation time show density temperature (DT) scaling  over a range of densities and temperatures  and the quality of data collapse improves if the negative virial states are excluded. Then we have analyzed the consequences of the DT scaling for the density dependence of fragility. We have found that although the kinetic fragility is not independent of density but the density dependence of the kinetic fragility is weak. We have shown that this weak density dependence can be understood in terms of DT scaling and deviations from DT scaling at low densities. We have further shown that the configurational entropy and hence the Adam-Gibbs (AG) relation exhibits DT scaling at high densities and consequently, the activation free energy in the AG relation show a power law dependence on density. The scaling exponent computed from DT scaling of  AG relation is consistent with the values obtained previously by other methods.


\end{document}